\documentclass[aps,prb,twocolumn,showpacs,showkeys]{revtex4} 
\usepackage[dvips]{graphicx}
\usepackage{bm}
\usepackage{simplewick}

\newcommand{\en}{\varepsilon_n} 
\newcommand{\ten}{\widetilde{\varepsilon}_n} 
\newcommand{\om}{\omega_m} 
\newcommand{\Om}{\Omega_m}

\newcommand{\ui}{{\rm i}} 
\newcommand{\bmq}{{\bm q}}
\newcommand{\bmk}{{\bm k}}
\newcommand{\bmK}{{\bm K}}
\newcommand{\bmp}{{\bm p}}
\newcommand{\bmP}{{\bm P}} 
\newcommand{\bmQ}{{\bm Q}} 

\newcommand{\rme}{{\rm e}}

\begin{document}

\title{
Probing the $d_{x^2-y^2}$-wave Pomeranchuk instability by ultrasound
}

\author{
Hiroto Adachi
}\altaffiliation{Present address: Institute for Materials Research,
Tohoku University, Sendai 980-8577, Japan}
\email[]{hiroto.adachi@imr.tohoku.ac.jp}
\author{ Manfred Sigrist} 
\affiliation{
Institute for Theoretical Physics, ETH Z\"{u}rich, Z\"{u}rich 8093, Switzerland} 

\date{\today}

\begin{abstract} 
Selection rules of ultrasound attenuation and sound velocity renormalization 
are analyzed in view of their potential application to identify 
Pomeranchuk instabilities (electronic nematic phase). 
It is shown that the transverse sound attenuation along $[110]$ direction 
is enhanced by the Fermi surface fluctuations near a $d_{x^2-y^2}$-wave 
Pomeranchuk instability, while the attenuation along $[100]$ direction 
remains unaffected. 
Moreover the fluctuation regime above the instability is analyzed 
by means of a self-consistent renormalization scheme. The results could be 
applied directly to Sr$_3$Ru$_2$O$_7$ which is a potential candidate for 
a Pomeranchuk instability at its metamagnetic transition in strong magnetic fields. 
\end{abstract}


\maketitle

\section{Introduction}  

In 1958, Pomeranchuk~\cite{Pomeranchuk} considered stability conditions for 
normal {\it isotropic} Fermi liquids and argued that, 
besides a uniform (spin-dependent) deformation of the Fermi surface, 
i.e., an itinerant ferromagnetic instability, 
residual quasiparticle interactions can induce a non-uniform deformation 
of the Fermi surface. 
Since then, an inhomogeneous deformation of a Fermi surface caused by 
quasiparticle interactions is frequently referred to as 
``Pomeranchuk instability (PI)'' It is sometimes called ``nematic'' electron 
liquid~\cite{Oganesyan01} because such a Fermi surface deformation can be 
characterized by a director order parameter which is used to describe 
a nematic phase in conventional liquid crystals. 
The nematic order introduces an anisotropy in the otherwise 
isotropic physical quantities, and 
the strong anisotropy of the longitudinal resistivity observed 
in a clean two-dimensional electron system under high magnetic fields 
was interpreted as a signal of the development of the nematic order.~\cite{Lilly99} 

In lattice systems the term PI is commonly used to refer to a 
Fermi surface deformation which breaks the symmetry of the underlying lattice. 
In case of two-dimensional square lattices, 
it was shown~\cite{Yamase00,Halboth,Hankevych,Valenzuela} that 
a PI with $d_{x^2-y^2}$ symmetry can be realized near van Hove fillings, 
which then reduces the $C_{4}$ symmetry of the original Fermi surface to $C_{2}$ 
symmetry. These findings have stimulated a number of further theoretical 
works~\cite{Honerkamp02,Frigeri02,Kee03a,Kee03b,
Metzner03,Arne03,Yamase04,Khavkine,Kee05,Yamase05,
DellAnna06,DellAnna07,Jakubczyk08} 
on the nature of a $d_{x^2-y^2}$-wave PI. 
Recently this phase has attracted a renewed interest in several fields of 
solid state physics. 
The occurrence of a $d_{x^2-y^2}$-wave PI is hotly debated 
as a possible explanation for the curious phase found in the 
bilayer ruthenate Sr$_3$Ru$_2$O$_7$ under strong magnetic 
field.~\cite{Grigera04,Borzi07} 
Also a $d_{x^2-y^2}$-wave PI may provide an explanation for 
the anisotropy in the magnetic excitation spectrum observed 
in cuprate superconductors,~\cite{Hinkov04,Hinkov08} 
and a similar argument is ongoing in the context of new iron-pnictide 
superconductors.~\cite{Xu08,Zhai09} 
An analogous state in the spin channel was also proposed~\cite{Hirsch90,Wu04,Wu07} 
and used to explain 
the so-called hidden-order phase found in URu$_2$Si$_2$~\cite{Varma05}. 

Despite these extensive discussions, however, 
an identification of such a $d_{x^2-y^2}$-wave PI is not an easy task. 
The problem lies in the difficulty in finding an experimental probe which 
couples to the order parameter of the $d_{x^2-y^2}$-wave PI. 
Unlike a magnetic transition where we can 
detect the magnetic susceptibility using neutron scattering 
experiment, it is hard to detect the corresponding susceptibility 
for a $d_{x^2-y^2}$-wave PI which occurs in the charge degrees of freedom. 
In Ref.~\onlinecite{Doh07}, it was suggested that a spatial pattern of the local 
density of states provides a direct probe of the $d_{x^2-y^2}$-wave PI. 
It is however applicable only through surface probes, a scanning 
tunneling microscopy, which is easily influenced by surface conditions. 
Therefore, it is desirable to have 
another way to identify the 
$d_{x^2-y^2}$-wave PI based on the {\it bulk} properties of a sample. 

From the point of view that the characteristic anisotropy of a 
$d_{x^2-y^2}$-wave PI can be easily wiped out in a macroscopic scale 
by a formation of domains between two degenerate ground states 
[FIG.~\ref{Fig:orderpara01} (b) and (c)],~\cite{Grigera04,Doh07} 
it had better concentrate on the fluctuation effects above the transition 
temperature~\cite{Emery76,Adachi01} 
rather than concentrate on the phenomena deep inside of the ordered state. 
The basic idea presented in this paper is that 
we can use phonons for the detection of the $d_{x^2-y^2}$-wave PI. 
This idea is based on the fact that phonons not only have a coupling to the order 
parameter of a $d_{x^2-y^2}$-wave PI through electron-phonon interactions, 
but also have a propagation-direction-selective coupling. 
By combining this remarkable property of phonons with a well known fact that 
damping of phonons (sound attenuation) exhibits 
a remarkable divergence near a second order phase transitions,~\cite{Landau54} 
we can pick up information on the $d_{x^2-y^2}$-wave PI. 
In other words, the fluctuation sound attenuation caused by a $d_{x^2-y^2}$-wave PI 
can be regarded as a direct probe of the unconventional 
charge susceptibility characterizing this ordered state. 

In this paper, we investigate the behavior of the fluctuation enhancement of the 
sound attenuation caused by a $d_{x^2-y^2}$-wave PI. 
For this purpose, we use two different approaches. 
The first one is a phenomenological argument using 
a Gaussian Ginzburg-Landau-Wilson action. 
The second one is a microscopic analysis starting from a quasi-two-dimensional 
Hubbard model with on-site and nearest-neighbor repulsions, which is 
supplemented by the self-consistent renormalization (SCR) theory~\cite{Moriya85} 
to treat the fluctuation effects. 
Both approaches are shown to give the same result that 
there is a selection rule in the fluctuation enhancement of the sound attenuation 
and sound velocity softening near a $d_{x^2-y^2}$-wave PI, 
depending on the propagation directions and polarizations. 
This leads to a conclusion that we can utilize these properties to detect the 
$d_{x^2-y^2}$-wave PI with the aid of ultrasound measurements. 

The plan of this paper is as follows. 
In Sec.~II, we present a phenomenological argument to see how 
the Fermi surface fluctuations near a $d_{x^2-y^2}$-wave PI 
manifest themselves in transverse sound attenuation. 
The main message of this paper is essentially given in this section. 
In Sec.~III, we present a microscopic analysis of a $d_{x^2-y^2}$-wave PI, 
and discuss the mean field phase diagram as well as the effects of fluctuations. 
Besides, we present a microscopic calculation of the fluctuation sound attenuation, 
which reinforces the result of the phenomenological argument in Sec.~II. 
Finally in Sec.~IV, we summarize our result and give a discussion on 
the application of our result to the bilayer ruthenate Sr$_3$Ru$_2$O$_7$. 
We use the unit $k_B=\hbar=1$ throughout this paper. 

\section{Phenomenological approach} 

\begin{figure}[t] 
  \begin{center}
    \scalebox{0.7}[0.7]{\includegraphics{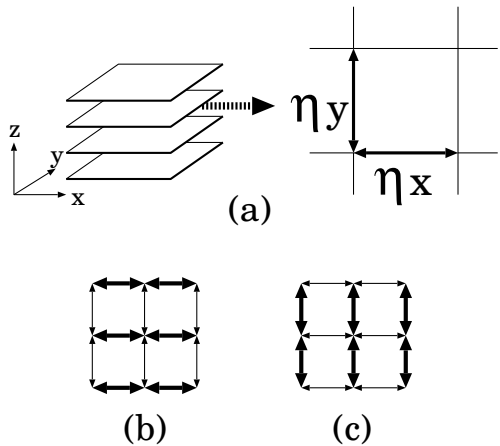}} 
  \end{center}
\caption{We consider a quasi two-dimensional system composed of 
a stack of two-dimensional square lattices. 
The order parameter of a $d_{x^2-y^2}$-wave PI 
considered in this work is defined by Eq.~(\ref{Eq:etaDEF}) 
where $\eta_x$ ($\eta_y$) represents the strength of the bond 
along $x$ ($y$) direction as shown in FIG.~\ref{Fig:orderpara01} (a). 
The two states in FIG.~\ref{Fig:orderpara01} (b) and (c) are degenerate 
where the thickness of arrows indicates the strength of the bond.} 
\label{Fig:orderpara01}
\end{figure}
 
In this section, we present a phenomenological argument to see how 
the Fermi surface fluctuations near a $d_{x^2-y^2}$-wave PI 
manifest themselves in phonon properties. 
We pay special attention to the propagation-direction resolved 
sound attenuation. 
Because we consider the fluctuation effects {\it above} the transition point 
and mainly focus on the {\it transverse} sound attenuation, 
our argument is essentially different from that given in 
Ref.~\onlinecite{Kee03a} where the {\it longitudinal }sound propagation 
{\it below} the transition point was discussed. 

The order parameter of a $d_{x^2-y^2}$-wave PI~\cite{Yamase00,Halboth} 
which is considered in this work is defined by [see also Eq.~(\ref{Eq:HV03})] 
\begin{eqnarray}
\eta &=& \frac{1}{2} \left( \eta_x - \eta_y \right), 
\label{Eq:etaDEF}
\end{eqnarray}
where $\eta_{x} (\eta_{y})$ denote the bond strength 
along $x$ ($y$) direction (Fig.\ref{Fig:orderpara01}) 
of the underlying square lattice. 
The order parameter $\eta$ is odd under 
the permutation $x \leftrightarrow y$. 
In the presence of nonzero $\eta$, the bond strength 
along $x$ and $y$ directions are different, hence it deforms the lattice 
and affects the phonon properties.

Up to the lowest order in $\eta$, 
the mode-coupling term $\delta F_{\rm GL}$ 
between lattice (phonon) and the PI order parameter 
is given by 
\begin{eqnarray} 
  \delta F_{\rm GL}({\bm U},\eta) &=& \kappa \sum_{{\bm Q}=(\bmq,Q_z)} 
  \Big[  u_{xx} ({\bm Q})- u_{yy} ({\bm Q}) \Big] \eta(-{\bm Q}), 
  \label{Eq:dFGL1a} 
\end{eqnarray}
where $\bmQ= (\bmq,Q_z)$ is a three-dimensional 
wavevector with the in-plane component $\bmq$ and the out-of-plane 
component $Q_z$, 
$\kappa$ is the coupling constant, 
$u_{xx}-u_{yy} = {\rm i}q_x u_x -{\rm i}q_y u_y$ 
with the lattice displacement field ${\bm U}= ({\bm u},U_z)$. 
The above form of the coupling is allowed from the symmetry reason 
because under the permutation $x \leftrightarrow y$ 
both $\eta$ and $u_{xx}-u_{yy}$ are odd, keeping the coupling invariant. 

We now study how phonons couple to the order parameter of a $d_{x^2-y^2}$-wave PI 
depending on their propagation direction and polarization. For the moment, 
we consider the case where the wavevector 
and the polarization vector of a sound wave lie within a conducting layer, 
i.e., $\bmQ= (\bmq,0)$ and ${\bm U}= (u_x,u_y,0)$, 
because interesting results come out in this case. 
Then, the longitudinal (transverse) phonon $u_L$ ($u_T$) can be written 
as $u_L= \widehat{q}_x u_x + \widehat{q}_y u_y $ 
($u_T= \widehat{q}_y u_x - \widehat{q}_x u_y $) 
with $\widehat{\bm q}= {\bm q}/|{\bm q}|$. 

Consider a sound wave propagating along 
$[100]$-direction (i.e., $\widehat{q}_y=0$). 
In this case, using the fact that $u_{xx}- u_{yy}= {\rm i}|{\bm q}| u_L$ when 
$\widehat{q}_y=0$, we can write $\delta F_{\rm GL}$ in Eq.~(\ref{Eq:dFGL1a}) as 
\begin{eqnarray}
  \delta F_{\rm GL}({\bm u},\eta) &=& \kappa \sum_{{\bm q}} 
  \Big( {\rm i} |{\bm q}| u_L ({\bm q}) \Big) \eta(-{\bm q}).  
  \label{Eq:dFGL1b}
\end{eqnarray}
On the other hand, when a sound wave is propagating along 
$[110]$-direction (i.e., $\widehat{q}_x=\widehat{q}_y$), 
the mode-coupling term (\ref{Eq:dFGL1a}) can be written as 
\begin{eqnarray}
  \delta F_{\rm GL}({\bm u},\eta) &=& \kappa \sum_{{\bm q}}  
  \Big( {\rm i} |{\bm q}| u_T ({\bm q}) \Big) 
  \eta(-{\bm q}) . 
  \label{Eq:dFGL1c} 
\end{eqnarray}
{Equations~(\ref{Eq:dFGL1b}) and (\ref{Eq:dFGL1c}) mean that, 
through the interaction (\ref{Eq:dFGL1a}), 
the longitudinal phonons couple to the $d_{x^2-y^2}$-wave PI only 
when they have their wavevector along [100]-direction, 
while transverse phonons does only when 
they have their wavevector along [110]-direction. }

Using these results [Eqs.~(\ref{Eq:dFGL1b}) and (\ref{Eq:dFGL1c})], 
we next study the effect of the fluctuations of $\eta$ 
on sound attenuation. 
Dynamical behaviors of sounds are described by the following 
action for phonons,~\cite{Kleinert} 
\begin{eqnarray}
  S_{\rm ph} ({\bm u}) &=& \frac{\rho_{\rm ion}}{2} \sum_{{\bm q},\om} 
  \Big( u_L^*({\bm q},\om) {K}_{L}({\bm q},{\rm i}\om) 
  u_L ({\bm q},\om) \nonumber \\
  && + \; {u}^*_T({\bm q},\om) 
  {K}_{T}({\bm q},{\rm i}\om) 
  {u}_T ({\bm q}, \om)\Big), 
  \label{Eq:Sph}
\end{eqnarray}
where 
the unrenormalized kernel has the form 
${K}^{(0)}_{\nu}({\bm q},{\rm i}\om) =  
  \om^2 + (s^{(0)}_\nu)^2 \; q^2  $ 
with $\nu$ being the polarization index $L$ or $T$. 
Here, 
$s^{(0)}_L$ and $s^{(0)}_T$ the bare longitudinal and transverse sound velocities, 
$\rho_{\rm ion}$ the mass density of ions, 
and $\om = 2 \pi T m$ is the bosonic Matsubara frequency. 
The information on phonon dynamics is extracted by studying 
the retarded quantity 
$K^{R}_{\nu}({\bm q},\omega)= K_{\nu}({\bm q},{\rm i}\om \to \omega + \ui 0_+)$. 
In the presence of itinerant electrons, 
excitations of particle-hole pairs yield an 
imaginary part in $K^{R}_{\nu}$ of the form~\cite{AGD} 
${\rm Im} K^{R}_{\nu}({\bm q},\omega)= -\omega/\tau_{{\rm ph}, \nu}({\bm q})$ 
for small $\omega$. 
This results in a finite phonon lifetime 
$\tau_{{\rm ph},\nu}({\bm q})$. 
The sound attenuation $\alpha_{\nu}({\bm q})$ is related to the 
phonon lifetime as $\alpha_{\nu}({\bm q})= 1/(s_\nu \tau_{{\rm ph},\nu})$, 
or equivalently, 
\begin{eqnarray}
\alpha_{\nu}({\bm q}) = 
-\frac{1}{s_\nu \omega}{\rm Im} K^{R}_{\nu}({\bm q},\omega). 
\end{eqnarray} 

The phonon propagator $[K^{R}_{\nu}({\bm q},\omega)]^{-1}$ is modified 
in the presence of $\eta$ fluctuations. When we consider the Gaussian fluctuation 
region for simplicity, the corresponding action is given by 
\begin{eqnarray} 
  S_{\rm GL}(\eta) &=& 
  \sum_{{\bm q},\om}  \eta^*({\bm q},\om) 
  \left[ \frac{1}{\chi_d({\bm q},{\rm i}\om)} \right]
  \eta({\bm q},\om),  
\end{eqnarray} 
where 
\begin{eqnarray} 
  \frac{1}{\chi_d({\bm q},{\rm i}\om)} &=& 
 \mu_0 + \xi_{0}^2 {\bm q}^2 
 + \frac{|\om|}{\Gamma_{d}(\bmq)}, 
 \label{Eq:chid01}
\end{eqnarray} 
is the $d$-wave density correlation function, 
$\mu_0 \propto T- T_{c0}$ measures the distance from the mean field 
transition temperature $T_{c0}$, 
$\xi_0$ and $\Gamma_{d}(\bmq)= w_0 |{\bm q}| $ are the bare correlation length 
and the damping rate of fluctuations, respectively  
[for the microscopic definition, 
see Eqs.~(\ref{Eq:xi0DEF}) and (\ref{Eq:w0DEF})]. 

After integrating out the $d_{x^2-y^2}$-wave PI order parameter $\eta$ 
by performing the Gaussian integral in 
$S_{\rm GL}(\eta)+ \delta S_{\rm GL}({\bm u},\eta)$ 
[where $\delta S_{\rm GL}({\bm u},\eta)= 
\int d \tau \; \delta F_{\rm GL}({\bm u},\eta)$], 
we can show that the couplings [Eqs.~(\ref{Eq:dFGL1b}) and (\ref{Eq:dFGL1c})] 
give an additional contribution 
\begin{eqnarray}
  \delta K_\nu({\bm q},{\rm i}\om) &=&  
  - \frac{\kappa^2 {\bm q}^2 }{4 \rho_{\rm ion}} \chi_d ({\bm q},{\rm i}\om) 
\end{eqnarray}
to $K_\nu({\bm q},{\rm i}\om)$. 
This leads to the divergent sound attenuation   
$\alpha_{\nu} ({\bm q}) \propto \frac{1}{\omega} 
{\rm Im} \chi_d^R ({\bm q},\omega)$. 
Recalling that the in-plane longitudinal phonon (transverse phonon) couples 
to $\eta$ only when the wavevector is along $[100]$ direction 
($[110]$ direction), we obtain 
\begin{eqnarray}
  \alpha_{L100}  &\propto& \mu_0^{-2}, \label{Eq:alpL100} \\
  \alpha_{T110}  &\propto& \mu_0^{-2}, \label{Eq:alpT110} 
\end{eqnarray}
while there are no divergent behaviors for sounds along the other directions, 
\begin{eqnarray}
  \alpha_{L110} &:& \text{no divergence}, \label{Eq:alpL110} \\
  \alpha_{T100} &:& \text{no divergence}. \label{Eq:alpT100} 
\end{eqnarray}
Here, $L100$ means the longitudinal sound wave 
propagating along $[100]$ direction, etc. 

In addition to Eq.~(\ref{Eq:dFGL1a}) there is another relevant term, 
\begin{eqnarray}
  \delta F'_{\rm GL} ({\bm u},\eta) 
  &=& \kappa' \sum_{{\bm q},{\bm q'}} 
  \eta^*({\bm q}+{\bm q'}) \eta({\bm q'}) 
  \Big[ {\rm i} |{\bm q}| u_L ({\bm q}) \Big], 
  \label{Eq:dFGL2a} 
\end{eqnarray}
in which the longitudinal sound modes couple 
to the $d_{x^2-y^2}$-wave PI order parameter. 
As before, 
by integrating out $\eta$ in $S_{\rm GL}(\eta)+ \delta S'_{\rm GL}({\bm u},\eta)$ 
[where 
$\delta S'_{\rm GL}({\bm u},\eta)= \int d \tau \; \delta F'_{\rm GL}({\bm u},\eta)$] 
and assuming a three-dimensional behavior of the fluctuations, 
we can show [see also the paragraph containing Eq.~(\ref{Eq:ALterm01})] 
that this coupling gives the longitudinal sound attenuation 
the {\it same} divergent contribution as 
Eqs.~(\ref{Eq:alpL100}) and (\ref{Eq:alpT110}), 
\begin{eqnarray}
  \alpha_{L}  &\propto& \mu_0^{-2} 
  \qquad \text{(in all directions)}. 
  \label{Eq:alpLall}
\end{eqnarray}
Note that this latter result is the same as that obtained in 
Ref.~\onlinecite{Paulson} 
for fluctuation sound attenuation near a magnetic transition, 
because the coupling (\ref{Eq:dFGL2a}) 
and the coupling discussed in Ref.~\onlinecite{Paulson} 
[Eq.~(1) therein] have the same form. 

\begin{figure}[t] 
  \begin{center}
    \scalebox{0.6}[0.6]{\includegraphics{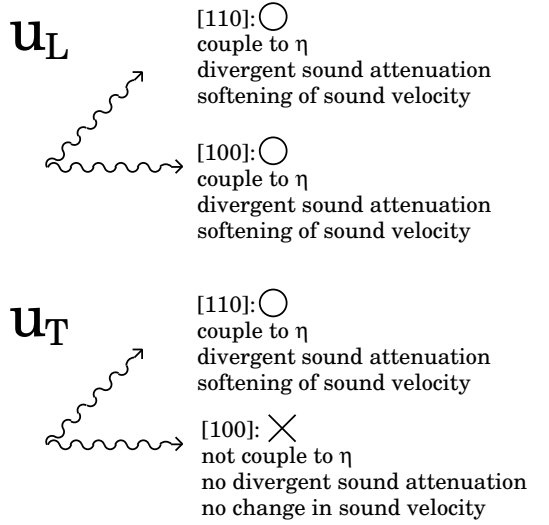}} 
  \end{center}
\caption{Summary of the results derived in Sec.~II. In the figure, $u_L (u_T)$ 
represents longitudinal (transverse) phonons with the wavy line being the 
direction of the wavevectors.} 
\label{Fig:uLuT}
\end{figure}

Before proceeding to the summary of this section, 
we comment on the behavior of the sound velocity. The anomaly 
in the sound attenuation is intimately related to the 
softening of the sound velocity. The renormalization of the sound 
velocity due to the fluctuation of $\eta$ is given by 
\begin{eqnarray}
s_\nu &=& 
s^{(0)}_\nu \sqrt{1+  \big(s^{(0)}_\nu q \big)^{-2} 
{\rm Re} \; \delta K^R_\nu({\bm q},\omega  \to 0)}. 
\end{eqnarray}
This means that  the divergence in the sound attenuation is accompanied by 
the reduction in the sound velocity (sound mode softening).

The results obtained in this section is summarized in FIG.~\ref{Fig:uLuT}. 
Eqs.~(\ref{Eq:alpL100})-(\ref{Eq:alpT100}) and (\ref{Eq:alpLall}) 
mean that longitudinal sound attenuation always show a divergent behavior 
on approaching a second order phase transition of a $d_{x^2-y^2}$-wave PI. 
On the other hand, in-plane transverse sound attenuation are divergent 
only when their wavevectors are along $[110]$-direction. 
Likewise, longitudinal sound velocities always show a critical sound mode 
softening 
near a $d_{x^2-y^2}$-wave PI, whereas in-plane transverse sound velocities 
show the softening 
only when their wavevectors are along $[110]$-direction. 
We therefore propose to detect the $d_{x^2-y^2}$-wave PI 
by using a measurement of propagation-direction resolved transverse 
sound attenuation and sound velocities. 

\section{Microscopic Analysis}  
In this section, we present a microscopic analysis of 
fluctuation sound attenuation near a $d_{x^2-y^2}$-wave PI 
in order to reinforce the phenomenological argument given in the previous section. 
Firstly, we perform a mean field analysis of a $d_{x^2-y^2}$-wave PI starting from 
a single-band Hubbard Hamiltonian with on-site and nearest-neighbor repulsions, 
and draw the mean field phase diagram. 
Next, we investigate the fluctuation effect above the transition temperature 
by employing the self-consistent renormalization formalism.~\cite{Moriya85} 
Finally, based on the same microscopic model, we present a diagrammatic calculation 
of the fluctuation sound attenuation above the transition temperature, which 
is shown to give the same conclusion derived within the more phenomenological approach of 
the previous section. 

\subsection{Mean field phase diagram} 
We start from a single-band Hubbard Hamiltonian 
on a two-dimensional square lattice, 
\begin{eqnarray}
  H &=& H_0 + H_{\rm imp} + H_{U}+ H_{V}, 
  \label{Eq:hamiltonian1} 
\end{eqnarray}
where 
\begin{eqnarray}
  H_U &=& U \sum_{{\bm r}_i} n_{\uparrow}({\bm r}_i)n_{\downarrow}({\bm r}_i) 
\end{eqnarray}
and 
\begin{eqnarray}
  H_V &=& V \sum_{\langle {\bm r}_i,{\bm r}_j \rangle} \sum_{\sigma,\sigma'} 
  n_{\sigma}({\bm r}_i)n_{\sigma'}({\bm r}_j) 
\end{eqnarray}
describe the on-site and nearest-neighbor repulsions. 
Here $n_{\sigma}({\bm r}_i)= c^{\dag}_{\sigma}({\bm r}_i) c_{\sigma}({\bm r}_i)$ 
is the number operator for electrons on a lattice site ${\bm r}_i$ with 
spin projection $\sigma=\pm 1$, and $\langle {\bm r}_i,{\bm r}_j \rangle$ 
means a bond between a nearest-neighbor lattice site ${\bm r}_i$ and ${\bm r}_j$. 

In this work we consider the effect of an external magnetic field 
$H_{\rm ext}$ perpendicular to the two-dimensional plane. 
Hence, the first term in Eq.~(\ref{Eq:hamiltonian1}) 
describes the kinetic energy plus the Zeeman energy, 
\begin{eqnarray}
  H_0 &=& \sum_{\bmp,\sigma} (\epsilon_{\bmp} - \epsilon_{\rm F}) 
  n_{\bmp,\sigma} -h \sum_{\bmp} (n_{{\bmp},\uparrow}- n_{{\bmp},\downarrow}), 
  \label{Eq:H_0}
\end{eqnarray}
where 
$n_{{\bmp},\sigma}= c^\dag_{\bmp,\sigma} c_{\bmp,\sigma}$ 
and $c_{\bmp,\sigma}= \frac{1}{\sqrt{N}} \sum_{{\bm r}_i} c_{\sigma}({\bm r}_i) 
{\rm e}^{-{\rm i}{\bmp}\cdot{\bm r}_i}$ 
with the number of lattice sites $N$. 
Here $\epsilon_{\bmp} = -2 t[\cos (p_x a) +\cos (p_y a)]
- 4 t' \cos (p_x a) \cos (p_y a)  
-2 t'' [ \cos (2p_x a)+\cos (2p_y a)] $  
is the single-particle dispersion 
with hopping amplitudes $-t, -t', -t''$ between nearest, 
next-nearest, and third nearest neighbors, 
$\epsilon_{\rm F}$ is the Fermi energy under the zero magnetic field, 
and $h= \mu_{\rm B} H_{\rm ext}$ with the Bohr magneton $\mu_{\rm B}$. 
Because we neglect the influence of magnetic field on orbital degrees of 
freedom (spin-orbit interaction, Landau diamagnetism), 
the magnetic-field effect is absorbed into 
the spin-dependent Fermi energy 
$\epsilon_{{\rm F},\sigma}(h)= \epsilon_{\rm F} + \sigma h$. 
Hereafter, Length and energy are measured in unit of the lattice spacing $a$ and 
the nearest hopping amplitude $t$.

The second term $H_{\rm imp}$ in Eq.~(\ref{Eq:hamiltonian1}) 
represents the spin-independent short-range isotropic ($s$-wave) 
impurity scattering, 
\begin{eqnarray} 
  H_{\rm imp} &=& 
  \sum_{{\bm r}_i,\sigma} u_{\rm imp}({\bm r}_i) n_{\sigma}({\bm r}_i), 
\end{eqnarray}
where the impurity potential $u_{\rm imp}$ obeys the Gaussian ensemble 
$\overline{u_{\rm imp}({\bm r}_i)}=0$, 
$\overline{u_{\rm imp}({\bm r}_i)u_{\rm imp}({\bm r}_j)}=n_{\rm imp}|u|^2 \delta_{i,j}$ 
with $n_{\rm imp}$ and $u$ being the impurity concentration and 
the strength of the impurity potential. 
We treat the impurity potential using Born approximation, and the 
impurity-averaged bare Green's function is given by 
\begin{eqnarray} 
G_{\sigma}^{(0)}(\bmp,\ui \en) &=& 
\frac{1}{\ui \ten - \xi_{\bmp,\sigma}}, 
\label{Eq:Green01}
\end{eqnarray}
where 
$\xi_{\bmk,\sigma}= \epsilon_\bmp - \epsilon_{{\rm F},\sigma}$, 
and 
$\ten= \en + \Gamma \, \text{sign}(\en)$ with the quasiparticle scattering rate 
$\Gamma$.

\begin{figure}[t] 
  \begin{center}
    \scalebox{0.8}[0.8]{\includegraphics{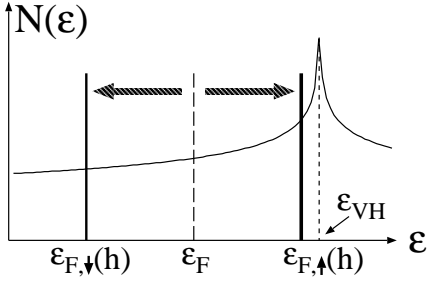}} 
  \end{center}
\caption{Schematic view of the density of states $N(\epsilon)$ 
considered in this paper. The density of states has a peak at van 
Hove energy $\epsilon_{\rm VH}=4(t'-t")$, and a spin-dependent Fermi energy 
$\epsilon_{{\rm F},\uparrow}$ touches $\epsilon_{\rm VH}$ under magnetic fields.
In the situation shown here, the up-spin is assigned as the ``active'' spin component.}
\label{Fig:VHcond}
\end{figure}

As discussed in Ref.~\onlinecite{Valenzuela}, Eq.~(\ref{Eq:hamiltonian1}) 
contains a phase with $d_{x^2-y^2}$-wave PI as a mean 
field solution. This phase appears when the Fermi energy coincides with 
van Hove energy~\cite{Halboth} $\epsilon_{\rm VH}= 4(t'-t'')$. 
In this paper we consider a situation where the zero-magnetic-field 
Fermi energy $\epsilon_{\rm F}$ does not satisfies the 
``van Hove condition'' ($\epsilon_{\rm F} \ne \epsilon_{\rm VH}$), 
but a moderately large external magnetic field $h$ tunes 
one of the spin-dependent Fermi energies 
$\epsilon_{{\rm F},\sigma^*}(h)$ 
to a van Hove condition. 
In this situation, the spin component $\sigma^*$ satisfying 
$\epsilon_{{\rm F},\sigma^*}(h) = \epsilon_{\rm VH}$ 
is relevant to the occurrence of the PI, and 
we hereafter consider only the ``active'' spin component $\sigma^*$. 
The antiferromagnetic correlations described by the on-site repulsion $H_U$ 
are suppressed because of the polarization under the strong external magnetic field. 
Although the ferromagnetic state could compete with the PI, we discuss the case 
where the PI is stabilized by choosing the parameters $V \approx U/2$.

We first make a mean field decoupling by taking the two contractions, 
\begin{eqnarray}
  H_V &=& V \sum_{\langle {\bm r}_i,{\bm r}_j \rangle} \sum_{\sigma,\sigma'} 
  \contraction[2ex]{}{c}{^{\dag}_{\sigma}({\bm r}_i) c_{\sigma}({\bm r}_i) 
    c^{\dag}_{\sigma'}({\bm r}_j) } {c}
  \contraction[3ex]{c^{\dag}_{\sigma}({\bm r}_i)} {c}{_{\sigma}({\bm r}_i)} {c} 
  c^{\dag}_{\sigma}({\bm r}_i) c_{\sigma}({\bm r}_i) 
  c^{\dag}_{\sigma'}({\bm r}_j) c
  _{\sigma'}({\bm r}_j), 
  \label{Eq:HV01} 
\end{eqnarray}
one by one. This yields the following mean field Hamiltonian, 
\begin{eqnarray}
  H^{\rm MF}_V &=& 
  -\frac{1}{2} \sum_{ {\bm r}_i }
  \sum_{ {\bm \delta}= \pm \hat{\bm x},\hat{\bm y} } 
  \eta_{\delta}({\bm r}_i) 
  c^{\dag}_{\sigma^*}({\bm r}_i) c_{\sigma^*}({\bm r}_i + {\bm \delta}), 
  \label{Eq:HV02} 
\end{eqnarray}
where 
$\eta_{\delta}({\bm r}_i) = 
V \langle c^{\dag}_{\sigma^*}({\bm r}_i + {\bm \delta} ) 
c_{\sigma^*}({\bm r}_i) \rangle, $
and we have picked up only 
the ``active'' spin component $\sigma^*$ satisfying the van Hove condition. 
We further assume that the parity of $\eta_{\delta}$ 
with respect to ${\bm \delta}$ is even (i.e., $\eta_{\delta}= \eta_{-\delta}$). 
Then, we have 
\begin{eqnarray}
  H^{\rm MF}_V &=& 
  -\frac{1}{2}\sum_{ {\bm r}_i } 
  \bigg[ \Big(\eta({\bm r}_i)+ \eta'({\bm r}_i) \Big) c^{\dag}_{\sigma^*}({\bm r}_i) 
  c_{\sigma^*}({\bm r}_i + \widehat{\bm x}) \nonumber \\ 
  &&+ \Big(-\eta({\bm r}_i)+ \eta'({\bm r}_i) \Big) c^{\dag}_{\sigma^*}({\bm r}_i) 
  c_{\sigma^*}({\bm r}_i + \widehat{\bm y} ) \bigg] \nonumber \\
  &&+ H.c. ,
  \label{Eq:HV03} 
\end{eqnarray}
where 
$\eta ({\bm r}_i) = \frac{1}{2}[ \eta_x ({\bm r}_i )- 
\eta_y ({\bm r}_i) ]$ and 
$\eta'({\bm r}_i) = \frac{1}{2} [\eta_x({\bm r}_i)+ \eta_y({\bm r}_i)]$. 
Because $\eta'$ possesses the four-fold symmetry of the 
underlying square lattice and can be regarded as a 
mass renormalization of the quasiparticles, we hereafter neglect it. 
On the other hand, a nonzero value of $\eta$ corresponds to the deformation of the 
Fermi surface which expands along the $p_x$ axis and shrinks along 
$p_y$ axis (or vice versa) and Eq.(\ref{Eq:HV03}) describes a 
$d_{x^2-y^2}$-wave PI. 
Going to the momentum representation, we have 
\begin{eqnarray}
  H^{\rm MF}_V &=& 
  - \sum_{ \bmq, \bmp } 
  d_\bmp \eta(\bmq) 
  c^{\dag}_{\bmp+\bmq,\sigma^*} c_{\bmp,\sigma^*}, 
  \label{Eq:HV04} 
\end{eqnarray}
where 
$d_\bmp= \cos p_x - \cos p_y$, and 
$\eta({\bm r}_i)= 
\sum_\bmq \eta(\bmq) e^{\ui \bmq {\bm r}_i}$. 

Several comments are in order. 
In a situation where the PI could be stabilized under the zero magnetic field, 
it would compete~\cite{Kee08} with the so-called 
$d$-density wave state, a state with $\eta(\bmq = \pm (\pi,\pi))$. 
Under a moderately large magnetic field, however, the $d$-density wave 
state is not expected to be competitive anymore since this state shows diamagnetic 
properties~\cite{Nersesyan1} suggesting that such a state is 
suppressed under sizable magnetic fields. 
Solving these problems is beyond our scope because 
it would require to include Landau diamagnetism such that the resulting analysis 
becomes considerably more complicated. 
Therefore in the following we assume that a uniform $\eta(\bmq=0)$ 
gives the ground state, and the momentum $\bmq$ appearing in 
Eq.~(\ref{Eq:HV04}) is small (i.e., $\bmq \ll 1$). 

\begin{figure}[t] 
  \begin{center}
    \scalebox{1.0}[1.0]{\includegraphics{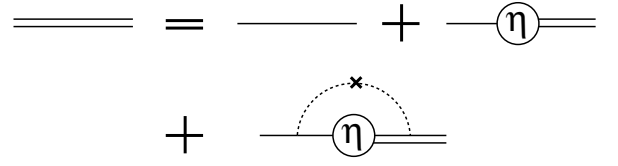}} 
  \end{center}
\caption{
The equation determining the impurity-averaged Green's function 
for the ``active'' spin component $\sigma^*$. 
The single solid line represent 
the impurity-averaged bare Green's function [Eq.~(\ref{Eq:Green01})]. 
A double solid line, and dotted line with a cross 
represent the impurity-averaged Green's function, and 
impurity scattering, respectively.}
\label{Fig:DysonEQ01}
\end{figure}

The equation determining the impurity-averaged Green's function 
for the ``active'' spin component $\sigma^*$ 
is diagrammatically shown in FIG.~\ref{Fig:DysonEQ01} where 
the single solid line represents 
the impurity-averaged bare Green's function [Eq.~(\ref{Eq:Green01})]. 
Hereafter, we neglect the last diagram in FIG.~\ref{Fig:DysonEQ01} 
assuming that the impurity potential is sufficiently weak. 
Due to the symmetry difference between $\eta$ and the impurity potential, 
this approximation is exact up to the linear order in $\eta$. 
Physically this means that the impurity effect on the 
second order transition line is rigorously treated while 
the first order transition line may slightly deviate from the 
exact result. We note here that even with this 
simplification, the required full numerical calculation is  rather involved. 
With this approximation, the equation for the impurity-averaged 
Green's function is solved as 
\begin{eqnarray} 
  G_{\sigma^*}(\bmp,\ui \en) &=& 
  \frac{1}{[G_{\sigma^*}^{(0)}(\bmp,\ui \en)]^{-1} + d_\bmp \eta}, 
  \label{Eq:DysonEQ01}
\end{eqnarray}
where the order parameter $\eta$ is determined by the following 
self-consistent equation 
\begin{eqnarray}
  \eta &=& V \int_{\bmp} d_\bmp f_\Gamma (\xi_{\bmp,\sigma^*} - d_\bmp \eta). 
  \label{Eq:selfconsist01}
\end{eqnarray}
Here we have introduced a shorthand notation 
$\int_\bmp = \frac{1}{N} \sum_\bmp$, and the quantity 
\begin{eqnarray}
f_\Gamma(\xi_{\bmp,\sigma^*}) 
  &=& T \sum_{\en} G^{(0)}_{\sigma^*} (\bmp,\ui \en) 
  \label{Eq:FDfunc}
\end{eqnarray}
may be considered as the impurity-averaged Fermi distribution function. 
The corresponding mean field free energy is given by 
\begin{eqnarray}
  \frac{F_{\rm MF}}{N} &=& 
  \frac{\eta^2}{2V} 
  - T \sum_{\en} \int_{\bmp} 
  \ln \frac{|G^{(0)}_{\sigma^*} (\bmp,\ui \en)|^2} 
  {|G_{\sigma^*} (\bmp,\ui \en)|^2 }. 
  \label{Eq:FMF01}
\end{eqnarray}

In the pure limit ($\Gamma \to 0$), the function $f_\Gamma$ 
becomes equal to the Fermi distribution function $f_0(x)= (1+ e^{(x)/T})^{-1}$ 
due to the identity 
$  T \sum_{\en} G^{(0)}_{\sigma^*} (\bmp,\ui \en) 
  = f_0(\xi_{\bmp,\sigma^*})$. 
This would simplify the self-consistent equation (\ref{Eq:selfconsist01}), 
and the free energy (\ref{Eq:FMF01}) would be reduced to a simpler form 
  $\frac{F_{\rm MF}}{N} = 
  \frac{\eta^2}{2V} 
  - T \int_{\bmp} 
  \ln | ( 1+ e^{-(\xi_\bmp- d_\bmp \eta)/T} )/(
  {1+ e^{-\xi_\bmp/T}} )|$. 
However because we consider the effect of non-magnetic impurities in this work, 
we need to use Eqs.~(\ref{Eq:selfconsist01}) and (\ref{Eq:FMF01}).

The mean field phase diagram can be drawn by making a Landau expansion~\cite{Yamase05} 
of the free energy (\ref{Eq:FMF01}) as 
\begin{eqnarray}
  \frac{F_{\rm MF}}{N} &\to& \frac{F_{\rm GL}}{N}
  = 
  \left( \frac{a_{\rm GL}}{2} \right) \eta^2 +  
\left( \frac{b_{\rm GL}}{4} \right) \eta^4 \cdots, 
  \label{Eq:GLcoeff01}
\end{eqnarray}
where the coefficients $a_{\rm GL}$ and $b_{\rm GL}$ are given by 
\begin{eqnarray}
  a_{\rm GL} &=& \frac{1}{V}-  
  \int_\bmp d^2_\bmp 
  \Big( -f'_\Gamma(\xi_{\bmp,\sigma^*}) \Big),   \label{Eq:GLcoeff02}\\ 
  b_{\rm GL} &=& 
  \frac{1}{3 !} \int_\bmp d^4_\bmp  
  \Big( f'''_\Gamma(\xi_{\bmp,\sigma^*}) \Big) ,   \label{Eq:GLcoeff03}
\end{eqnarray}  
with a convention 
$f'_\Gamma(\xi_{\bmp,\sigma})=  \frac{\partial } {\partial \xi_{\bmp,\sigma}} 
f_\Gamma(\xi_{\bmp,\sigma}) $, etc. 

A condition $a_{\rm GL}=0$ defines a second order transition 
line $T_{c0}(h)$ in the mean field approximation, if 
the coefficient of the quartic term of the GL free energy ($b_{\rm GL}$) is 
positive. In case of a negative quartic term the transition becomes 
of the first order, and the first order transition line $T^*_{c}(h)$ 
is determined by the conditions $F_{\rm MF}(\eta)=0$ and 
$\partial F_{\rm MF}(\eta)/\partial \eta=0$ for a nonzero $\eta$ 
where $F_{\rm MF}$ is defined in Eq.~(\ref{Eq:FMF01}). 

\begin{figure}[t] 
  \scalebox{0.5}[0.5]{\includegraphics{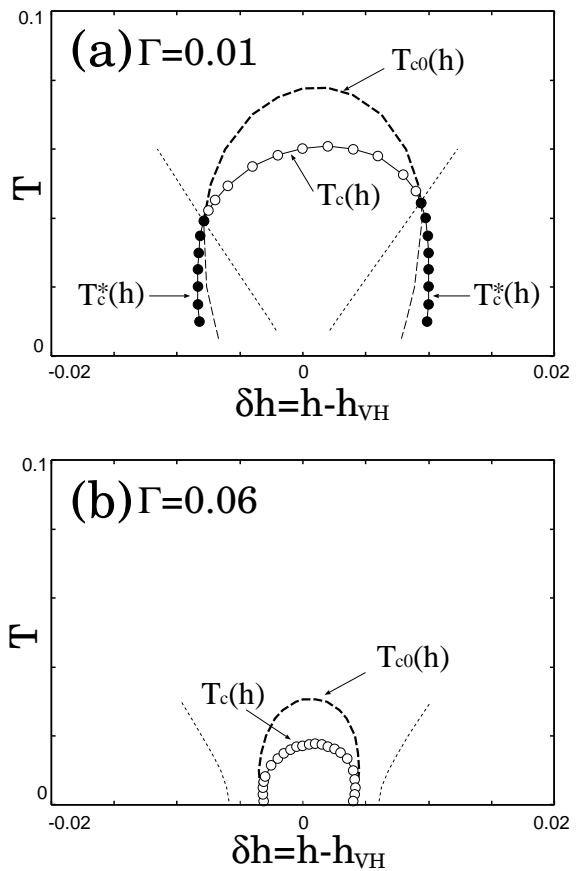}} 
\caption{Phase diagrams obtained from 
Eq.~(\ref{Eq:hamiltonian1}) for two different strengths 
of the impurity scattering, $\Gamma=0.01$ [(a)] 
and $\Gamma=0.06$ [(b)]. Note that length and energy 
are measured in unit of the lattice spacing $a$ and 
the nearest hopping amplitude $t$. 
We use $U=3.0$ and $V=1.5$ for 
electron repulsions, and $t'=0.4$ and $t''=0.2$ for hopping. 
Two (thick and thin) dashed lines are defined by $a_{\rm GL}=0$ 
where the thick dashed line represents a mean field second 
order phase transition line $T_{c0}(h)$ while the thin dashed line 
is an unphysical one because it is replaced by the following first order transition. 
A line with full circles show a first order transition line 
$T^*_{c}(h)$ defined by 
$F_{\rm MF}(\eta)=0$ and $\partial F_{\rm MF}(\eta)/\partial \eta=0$. 
A line with open circles represents the fluctuation renormalized 
second order transition line $T_c(h)$. 
A dotted line represents a temperature below which 
the quartic term $b_{\rm GL}$ [Eq.~(\ref{Eq:GLcoeff03})] of the 
Ginzburg-Landau free energy becomes negative. } 
\label{Fig:MFPD}
\end{figure} 

We calculate the mean field phase diagram by using a square mesh of $500 \times 500$
in the Brillouin zone. 
In FIG.~\ref{Fig:MFPD} (a), we show the calculated mean field phase diagram 
for a moderately clean sample ($\Gamma=0.01$). 
Not surprisingly, the phase diagram in this clean case 
is quite similar to that already obtained in the pure limit ($\Gamma=0$) 
in the previous work,~\cite{Yamase07} since as will be discussed below 
Eq.~(\ref{Eq:HV06}) in the next section, our model is similar to that used in 
Ref.~\onlinecite{Yamase07} except for the difference in a spin mixing. 
The important feature in the phase diagram is that 
the transition is of second order at higher temperatures while 
at lower temperatures a first order transition occurs. 
In the figure, we used a non-zero third-nearest neighbor hopping $t''=0.2$. 
This is because when we use smaller values of $t''$, a finite 
momentum state $\eta(\bmq \ne 0)$ is stabilized due to a nesting condition. 

Next let us see how a slight increase in the impurity scattering 
modifies the mean field phase diagram.~\cite{Ho08} 
In FIG.~\ref{Fig:MFPD} (b), the mean field phase diagram 
for a dirtier case ($\Gamma=0.06$) is shown. 
Firstly, we see that the PI is easily suppressed by the 
increased impurity scattering. This is understandable because 
the order parameter of the PI and the impurity potential have 
different symmetry. 
Hence, the underlying mechanism is analogous to what happens 
to a disordered $d$-wave superconductor where Anderson's theorem is 
violated due to the unconventional nature of the order parameter. 
Secondly, the first order transition line at low temperatures 
disappears while the second order transition still survives. 
These two features concerning impurity effects are important when 
we discuss the relevance of the PI-scenario to the metamagnetic transition 
found in Sr$_3$Ru$_2$O$_7$ in Sec.~IV.

\subsection{Self-consistent Renormalization (SCR) treatment of the Fermi surface 
fluctuation} 
In this subsection, we study how the mean field result obtained 
in the previous subsection is modified by fluctuation effects. 
Following Refs. \onlinecite{Yamase04} and \onlinecite{DellAnna06}, we define the 
(thermal) $d$-wave density correlation function 
\begin{eqnarray}
  \chi_d(\bmq,\ui \om) &=& 
  \frac{1}{N} \int_0^{1/T} d \tau \; 
  \rme^{\ui \om \tau} 
  \langle n_d(\bmq,\tau) n_d(-\bmq,0) \rangle, 
 \label{Eq:chid02}
\end{eqnarray}
where 
  $n_d(\bmq) = 
  \sum_{\bmp} d_{\bmp} c^\dag_{\bmp+\bmq/2,\sigma^*} 
  c_{\bmp-\bmq/2,\sigma^*} $
is the $d$-wave density operator in the ``active'' spin component $\sigma^*$. 
In general, $\chi_d$ has the following structure 
\begin{eqnarray}
  1/ \chi_d &=& 1/\chi_d^{\rm irr} - V, 
  \label{Eq:BS}
\end{eqnarray}
where $\chi^{\rm irr}_d$ is the irreducible part 
of $\chi_d$. 
To proceed further to the detailed calculation of $\chi_d$, 
it is convenient to rewrite the interaction Hamiltonian $H_V$ 
in the momentum space, 
\begin{eqnarray}
  H_V &=& \frac{1}{2N} \sum_{\bmp_{1},\bmp_{2}, \bmp_{3}} \sum_{\sigma,\sigma'}
  V_{\bmp_{1}-\bmp_{2}} c^\dag_{\bmp_{1},\sigma} c_{\bmp_{2},\sigma} \nonumber \\
  && \hspace{2.5cm} \times c^\dag_{\bmp_{3},\sigma'} 
  c_{\bmp_{1}+\bmp_{3}-\bmp_{2},\sigma'}, 
  \label{Eq:HV05} 
\end{eqnarray}
where $V_{\bmp}= 2V(\cos p_x + \cos p_y)$. 
Now we decouple $V_{\bmp_{1} - \bmp_{2}}$ as 
$V_{\bmp_{1} - \bmp_{2}} = 
 V [ d_{\bmp_{1}} d_{\bmp_{2}} + s_{\bmp_{1}} s_{\bmp_{2}} 
+ p^{(+)}_{\bmp_{1}}p^{(+)}_{\bmp_{2}} + 
  p^{(-)}_{\bmp_{1}}p^{(-)}_{\bmp_{2}} ]$, 
where $d_{\bmp}= \cos k_x - \cos k_y$, $s_{\bmp}= \cos k_x + \cos k_y$, 
and $p^{(\pm)}_{\bmp}= \sin k_x \pm \sin k_y$. 
Because the four van Hove points in the Brillouin zone enhance the 
interaction with $d_{x^2-y^2}$ form factor~\cite{Halboth} $d_{\bmp}$, 
we hereafter consider only this channel. 
After changing the ordering of fermion operators and 
setting $\bmp_{1}=\bmp+ \bmq/2, 
\bmp_{2 }= \bmp'+ \bmq/2,
\bmp_{3}=\bmp'-  \bmq/2$, 
we have 
\begin{eqnarray}
  H_V &\to& H'_V 
  = - \frac{V}{2N} \sum_{\bmq} n_d(\bmq) n_d (-\bmq). 
  \label{Eq:HV06} 
\end{eqnarray}
This is essentially the same model as 
used in Refs.\onlinecite{Metzner03} and \onlinecite{Yamase07} except for the fact that 
the interaction $H'_V$ acts only among the active spin component $\sigma^*$. 

The so-called random phase approximation (RPA) for $\chi_d$ is obtained 
when we adopt the simplest building block 
\begin{eqnarray}
  \chi^{{\rm irr}}_{d,0}(\bmq,\ui \om) &=& 
  - T \sum_{\om} \int_{\bmp} 
  d_{\bmp}^2 G_{\sigma^*}^{(0)}(\bmp,\ui \en) \nonumber \\
  &&\hspace{0.1cm}\times  G_{\sigma^*}^{(0)}(\bmp+\bmq, \ui \en + \ui \om) 
  \label{Eq:chi_irr_d0}
\end{eqnarray}
as the irreducible $\chi^{\rm irr}_d$. 
Hereafter, we write the $d$-wave density correlation function 
in a dimensionless form by introducing 
$\widetilde{\chi}_d = \chi_d/\chi^{\rm irr}_{d,0}$. 
Then, the RPA $d$-wave density correlation function is given by 
\begin{eqnarray}
  1/\widetilde{\chi}_{d,{\rm RPA}}(\bmq,\ui \om) &=& 
  1 - V \chi^{\rm irr}_{d,0}(\bmq,\ui \om). 
\end{eqnarray}
Near the second order PI, the retarded function 
$\widetilde{\chi}^{R}_{d,{\rm RPA}}(\bmq,\omega)= 
\widetilde{\chi}_{d,{\rm RPA}}(\bmq,\ui \om \to \omega + \ui 0_+)$ 
takes the same form as Eq.~(\ref{Eq:chid01}), 
\begin{eqnarray}
  [\widetilde{\chi}^R_{d,{\rm RPA}} (\bmq,\omega)]^{-1} &=& 
  \mu_0 + \xi_0^2 q^2 
  - \ui \frac{\omega}{w_0(\widehat{\bmq}) |q| }. 
  \label{Eq:PRA01}
\end{eqnarray} 
where $\mu_0 = V a_{\rm GL} \propto T- T_{c0}$ with 
$a_{\rm GL}$ and $T_{c0}$ being the quadratic coefficient 
in Eq.~(\ref{Eq:GLcoeff02}) 
and the mean field transition temperature. 
Also, $\xi_0^2$ and $w_0$ are given by (see Appendix A) 
\begin{eqnarray}
  \xi_0^2 &=& 
  V \int_\bmp d^2_\bmp \Big[ 
  \frac{f{'''}_\Gamma(\xi_{\bmp,\sigma^*})}{12} 
  (v_x^2 +v_y^2) \nonumber \\ 
  && \hspace{1cm} + \frac{f{''}_\Gamma(\xi_{\bmp,\sigma^*}) }{4} 
  (M_x   +M_y) \Big], 
  \label{Eq:xi0DEF}\\
  \frac{1}{w_0(\widehat{\bmq})} 
  &=& 
  V \int_{\bmp} d^2_{\bmp} 
  \left( \frac{
      - f'_\Gamma(\xi_{\bmp,\sigma^*})  }
    { 2 \Gamma /q_{{\rm c}}  
    +\ui {\bm v}_{\bmp,\sigma^*} \cdot \widehat{\bmq}} 
  \right), 
  \label{Eq:w0DEF}
\end{eqnarray}
where $M_j= (1/2)\partial^2 \xi_\bmp /\partial k_j^2 $ for $j=x,y$ 
and $q_{{\rm c}}$ is the wavevector cutoff. 
We note that, as was already mentioned above Eq.~(\ref{Eq:DysonEQ01}), 
neglecting the impurity vertex corrections,~\cite{Fulde68} 
which would otherwise transform the dynamical behavior into a diffusive one,  
remains exact in the calculation of these quantities. 

\begin{figure}[t] 
\begin{center} 
  \scalebox{0.55}[0.55]{\includegraphics{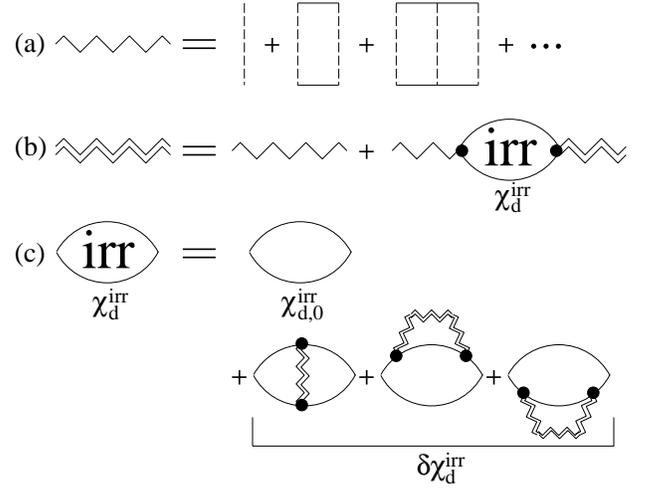}}
\end{center} 
\caption{Diagrams for the fluctuation propagator of $d_{x^2-y^2}$-wave 
PI [(a) and (b)] and the irreducible $d_{x^2-y^2}$-wave correlation 
function giving the mode coupling between fluctuations [(c)]. 
The solid line represents the impurity-averaged bare Green's function 
of quasiparticles, the dashed line the next nearest neighbor repulsion 
[Eq.~(\ref{Eq:HV05})].} 
\label{Fig:fluct_diag01}
\end{figure}

In the presence of interlayer coupling, the critical behavior 
is expected to be governed by three-dimensional fluctuations,~\cite{Hikami80} 
and we assume in this work an anisotropic three-dimensional (3D) 
behavior of the fluctuations. 
Microscopically, the three 
dimensionality is introduced by adding a $z$-axis dispersion 
$\delta \epsilon_{P_z}= - \delta t \cos(P_z)$ into Eq.~(\ref{Eq:H_0}). 
Instead of doing so, however, we here introduce the three dimensionality 
in a more phenomenological way, 
by replacing the two-dimensional wavevector $\bmq$ appearing in 
$\widetilde{\chi}_d(\bmq,\ui \om)$ with a three-dimensional wavevector 
$\bmQ' = (\bmq, Q_z/\gamma)$, 
namely, 
\begin{eqnarray}
\widetilde{\chi}_d(\bmq,\ui \om) &\to& \widetilde{\chi}_d(\bmQ',\ui \om), 
\label{Eq:Qaniso}
\end{eqnarray}
where $\gamma= \xi_{0}/\xi_{c0}$ is the anisotropy parameter 
given by the ratio of the in-plane correlation length to the 
out-of-plane correlation length. 
The above procedure is justified at least for a system with 
anisotropic ellipsoidal 3D Fermi surface. 
Here we would like to mention the recent proposal in the context of 
high-$T_c$ cuprates~\cite{Yamase09} that 
the interlayer-configuration of the nematic order becomes 
alternate pattern of $\eta > 0$ [FIG.~\ref{Fig:orderpara01} (b)] 
and $\eta < 0$ [FIG.~\ref{Fig:orderpara01} (c)]. 
We do not take into account such a possibility here, since
this configuration costs the elastic energy of the crystal lattice. 
We would like to point out that, 
while a minor change in the form of the propagator in Eq.~(\ref{Eq:PRA01}) is needed 
if the $d_{x^2-y^2}$-wave PI has such a configuration, 
the main result of this paper is not changed, at least qualitatively.

Now we consider the effect of mode coupling between the fluctuations. 
In the bosonic languages,~\cite{Hertz76,Millis93} this is given 
by the self-energy renormalization for the fluctuation propagator 
coming from the quartic term of the action. In the fermionic languages 
which we adopt in this work, the corresponding mode coupling 
is given by the three diagrams~\cite{Kawabata} shown in 
Fig.~\ref{Fig:fluct_diag01}, which give 
\begin{eqnarray}
  \widetilde{\chi}^{\rm irr}_d &=& \widetilde{\chi}^{\rm irr}_{d,0} 
  + \delta \widetilde{\chi}^{\rm irr}_{d}
  \label{Eq:chiIRR01},\\
  \delta \widetilde{\chi}^{\rm irr}_{d}  &=& -3 b_{\rm GL} \; V^2 \; T \sum_{\om} 
  \int \frac{\gamma d^3 Q'}{(2 \pi)^3} \; \widetilde{\chi}_d (\bmQ',\ui \om), 
  \label{Eq:chiIRR02}
\end{eqnarray}
where $b_{\rm GL}$ is the coefficient of the quartic term in the 
Ginzburg-Landau free energy 
given by Eq.~(\ref{Eq:GLcoeff01}). 
Substituting Eq.~(\ref{Eq:chiIRR01}) into Eq.~(\ref{Eq:BS}) and expanding it 
in terms of $\delta \chi^{\rm irr}_{d}$, we obtain 
\begin{eqnarray}
  1/\widetilde{\chi}_{d} &=& 
  1/\widetilde{\chi}_{d,{\rm RPA}} - \delta \widetilde{\chi}^{\rm irr}_{d}, 
  \label{Eq:SCR01}
\end{eqnarray}
where $\delta \widetilde{\chi}^{\rm irr}_{d}$ also depends on 
$\widetilde{\chi}_{d}$ through Eq.~(\ref{Eq:chiIRR02}).

As discussed in Ref.~\onlinecite{Millis93}, the SCR formalism is 
a self-consistent one-loop approximation for the mass renormalization 
\begin{eqnarray}
  \mu_{} &=& 1/\widetilde{\chi}_{d} (\bmQ'={\bm 0},\ui \om = 0). 
  \label{Eq:deltaSCR} 
\end{eqnarray}
Substituting Eq.~(\ref{Eq:deltaSCR}) into Eq.~(\ref{Eq:SCR01}) and 
making the contour deformation for the Matsubara summation, we obtain the 
following SCR equation to determine $\mu_{}$, 
\begin{eqnarray}
  \mu_{} &=& \mu_0 + \delta \mu, \\
  \delta \mu &=& 3 b_{\rm GL} V^2 \int_{-\infty}^\infty \frac{d\omega}{2 \pi}  
  \int_0^{Q'_{\rm c}} 
  \frac{\gamma d^3 Q'}{(2 \pi)^3} \coth \left( \frac{\omega}{2 T} \right) 
  \nonumber \\
  && \hspace{2cm} \times {\rm Im} \widetilde{\chi}^R_{d} (\bmQ',\omega), 
  \label{Eq:SCR02} 
\end{eqnarray}
where ${Q}'_{\rm c}$ is the  wavevector cutoff. 
The frequency integral in Eq.~(\ref{Eq:SCR02}) can be performed 
using the relation $\coth(x/2)= 1+ 2/(\rme^z -1)$ 
and $\int_0^\infty dx (\rme^x-1)^{-1}[x/(x^2+c^2)]= 
(1/2)[\ln(c/2 \pi)- \pi/c - \Psi(c/2 \pi)]$. 
Then, introducing the normalized wavevector $\widetilde{\bmQ}'= \xi_0 \bmQ'$, 
the final result can be written as 
\begin{eqnarray}
  \delta \mu &=& \frac{3 b_{\rm GL} V^2 \gamma}{2 \pi^2 \xi_0^3} 
  \int_0^{\widetilde{Q}'_{\rm c}} (\widetilde{Q}'^2 d \widetilde{Q}') 
  \left( \frac{T}{\mu_{ \widetilde{Q}' } } \right) \nonumber \\
  && \hspace{1cm} \times 2 Z_{\widetilde{Q}'} \Big( \ln Z_{\widetilde{Q}'} 
  -\frac{1}{2 Z_{\widetilde{Q}'}} - \Psi(Z_{\widetilde{Q}'}) \Big), 
  \label{Eq:SCR03} 
\end{eqnarray}
where $\Psi(Z)$ is the digamma function, 
$\mu_{\widetilde{Q}'}= \mu_{}+ \widetilde{Q}'^2$, 
$Z_{\widetilde{Q}'}= 
\mu_{\widetilde{Q}'} w_0 |\widetilde{Q}'|/(2 \pi T \xi_0 )$. 
When the condition $Z_{\widetilde{Q}'} \ll 1$ is satisfied, 
we can set $2Z_{\widetilde{Q}'} [\ln Z_{\widetilde{Q}'} -(1/2)Z_{\widetilde{Q}'} - 
\Psi(Z_{\widetilde{Q}'})] \to 1$ and 
the above expression reproduces the result of 
the Hartree approximation for the 
mass renormalization due to classical fluctuations,~\cite{Chaikin} 
which means that Eq.~(\ref{Eq:SCR03}) can describe both quantum and classical 
fluctuation regions. 

\begin{figure}[t] 
\begin{center}
  \scalebox{0.5}[0.5]{\includegraphics{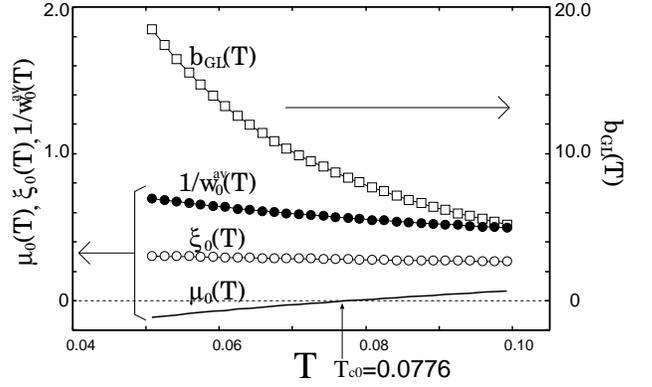}}
\end{center}
\caption{Temperature dependences of parameters characterizing 
the bare $d$-wave density correlation function; 
the mass term $\mu_0$, correlation length $\xi_0$, 
and damping coefficient $w_0$. 
The mode-coupling parameter $b_{\rm GL}$ is also plotted. 
The parameters used in this figure are the same 
as in FIG.~\ref{Fig:MFPD} (a). 
} 
\label{Fig:GLpara01}
\end{figure}

In FIG.~\ref{Fig:GLpara01}, we plot the temperature dependences 
of several parameters characterizing the bare $d$-wave density 
correlation function, the mass term $\mu_0$, correlation length $\xi_0$, 
and damping coefficient $w_0$, 
as well as the mode-coupling parameter $b_{\rm GL}$. 
Strictly speaking, $w_0(\widehat{\bmq})$ in Eq.~(\ref{Eq:w0DEF}) 
depends on $\widehat{\bmq}$ but we average them using two values 
at $\bmq= \widehat{\bm x}$ and 
$\bmq= (\widehat{\bm x}+\widehat{\bm y})/2$, i.e., 
$w_0(\bmq) \to w_0^{\rm av}= (1/2)\{ w_0(\widehat{\bm x})+ 
w_0[(\widehat{\bm x}+\widehat{\bm y})/2]\}$. 
From the figure, we can see the following three points. 
First, as was already discussed in the 
pure limit~\cite{Yamase07} ($\Gamma=0$), the bare mass term $\mu_0$ 
is as small as $\sim 0.1$ in a wide temperature region 
above the mean field transition temperature $T_{c0}$. 
Second, the correlation length is shorter than the lattice spacing 
$a=1$, which can be understood from Eqs.~(\ref{Eq:xi0DEF}) because 
the quasiparticle velocity $|{\bm v}_\bmp|$ is quite small 
near van Hove filling. Third, the mode coupling parameter $b_{\rm GL}$ 
is one order of magnitude larger than other parameters 
($\mu_0,\xi_0,1/w_0$) characterizing Gaussian fluctuations. 
All these features result in strong fluctuations near the $d_{x^2-y^2}$ PI. 

\begin{figure}[t] 
\begin{center}
  \scalebox{0.54}[0.54]{\includegraphics{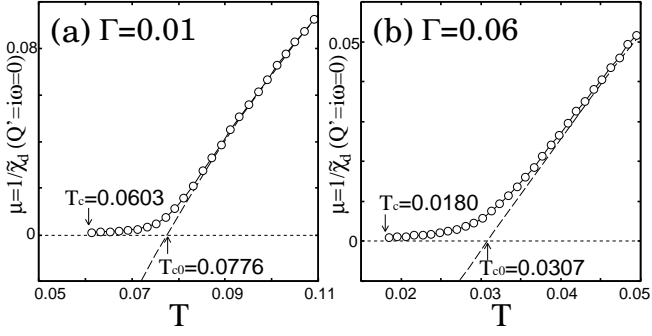}}
\end{center}
\caption{Temperature dependence of the renormalized 
mass $\mu = 1/\widetilde{\chi}_{d} (\bmQ'={\bm 0},\ui \om = 0)$ 
(open circles) and the bare mass 
$\mu_0 = 1/\widetilde{\chi}_{d,{\rm RPA}} (\bmQ'={\bm 0},\ui \om = 0)$ 
(dashed line) for $\Gamma=0.01$ [(a)] and $\Gamma=0.06 $ [(b)]. 
We use the wavevector cutoff ${Q}'_{\rm c}=0.0555$ and anisotropy parameter 
$\gamma=5$, and other parameters are the same as in FIG.~\ref{Fig:MFPD}. 
Two curves are calculated along a fixed magnetic field $h=h_{\rm VH}$ 
where the mean field transition is always of second order. 
The fluctuation-renormalized transition temperature $T_c$ is 
determined by a condition $\mu(T_c) < 10^{-3}$. 
}
\label{Fig:alpR-temp}
\end{figure}

In FIG.~\ref{Fig:alpR-temp}, 
we plot the temperature dependence of the renormalized 
mass 
$\mu = 1/\widetilde{\chi}_{d} (\bmQ'={\bm 0},\ui \om = 0)$ 
as well as the bare mass 
$\mu_0 = 1/\widetilde{\chi}_{d,{\rm RPA}} (\bmQ'={\bm 0},\ui \om = 0)$. 
As expected, $\mu(T)$ coincides with $\mu_0(T)$ 
well above the mean field transition temperature $T_{c0}$ 
but it deviates from the mean field result near and below $T_{c0}$.  
From the figure we determine the fluctuation-renormalized transition 
temperature $T_c$ by a condition $\mu(T_c) < 10^{-3}$. 
In FIG.~\ref{Fig:MFPD} we have also plotted $T_c(h)$ thus determined. 
We can see that the fluctuation region becomes smaller on approaching 
the first order transition line. 

\subsection{Fluctuation sound attenuation} 
We consider the electron-phonon interaction following the argument 
given by Walker, Smith and Samokhin,~\cite{Walker01} 
and begin with the tight-binding Hamiltonian 
\begin{eqnarray}
  H_{\rm el-ph} &=& \sum_{\bm \delta} 
  {\bm g}({\bm \delta}) \sum_{{\bm r}_i, \sigma} 
  \Big( {\bm u}({\bm r}_i) - {\bm u}({\bm r}_i + {\bm \delta} )\Big) \nonumber \\
  &\times&  \Big( c^{\dag}_{\sigma}({\bm r}_i) c_{\sigma}({\bm r}_i+ {\bm \delta}) 
  + c^{\dag}_{\sigma}({\bm r}_i+ {\bm \delta}) c_{\sigma}({\bm r}_i) 
  \Big), 
\end{eqnarray}
where ${\bm u}({\bm r}_i)$ is the displacement of the ion at ${\bm r}_i$, 
${\bm \delta}$ is the lattice vector, 
${\bm g}({\bm \delta}) = 
{\bf \nabla} t({\bm \delta})$ with the hopping amplitude $t({\bm \delta})$ 
between site ${\bm r}_i$ and ${\bm r}_i+ {\bm \delta}$. 
Following the procedure in Ref.~\onlinecite{Walker01} and neglecting umklapp 
processes, the interaction between electrons and a sound wave with wavevector 
$\bmK$ and polarization $\nu$ is given by 
\begin{eqnarray}
  H_{\rm el-ph} &=& \frac{1} {\sqrt{N}} 
  \sum_{\bmP,\bmK,\sigma} 
  F_{\{\widehat{\bmK},\nu\}}(\bmP) \nonumber \\ 
  && \hspace{1cm} \times \sqrt{\frac{\omega^{(0)}_{\bmK,\nu} }{2} } 
  B_{\bmK,\nu} c^\dag_{\bmP+\bmK,\sigma} c_{\bmP,\sigma}, 
  \label{Eq:H_el-ph}
\end{eqnarray}
where $B_{\bmK,\nu}= 
b_{\bmK,\nu} - b^\dag_{-\bmK,\nu}$ 
with the phonon annihilation operator $b_{\bmK,\nu}$, 
$\widehat{\bm e}_\nu$ is the polarization vector, and  
the electron-phonon vertex function $F_{\{\widehat{\bmK},\nu\}}(\bmP)$ 
is given by 
\begin{eqnarray} 
F_{\{\widehat{\bmK},\nu\}}(\bmP) &=& 
\frac{-2}{\sqrt{ \rho_{\rm ion} s_\nu^2 }} 
\sum_{\bm \delta} 
(\widehat{\bmK} \cdot \widehat{\bm \delta})
(\widehat{\bm e}_\nu \cdot {\bm g}({\bm \delta}) ) \nonumber \\
&& \hspace{2cm} \times \cos (\bmP \cdot {\bm \delta}). 
\label{Eq:Vertex_el-ph}
\end{eqnarray}
Here, all momenta denoted by capital letters are 
3D vectors. However, we consider below the case where the wavevector 
and the polarization vector of a sound wave lie within a conducting layer 
because interesting results come out in this case, 
and we set in the following 
$\bmP= (\bmp,0)$, $\bmK= (\bmk,0)$, etc. 
In this work we consider the electron-phonon coupling through 
at most the next nearest neighbor interactions. Assuming 
that ${\bm g}({\bm \delta})$ is proportional to $\widehat{\bm \delta}$, 
we make an expansion 
${\bm g}({\bm \delta}) = g_{\rm n.n.} \widehat{\bm \delta}_{\rm n.n.}
+ g_{\rm n.n.n.} \widehat{\bm \delta}_{\rm n.n.n.}$ where 
$g_{\rm n.n.} (g_{\rm n.n.n.})$ is the coupling constant through 
the nearest (next nearest) neighbor interactions. 

Then, for transverse phonons propagating along the $[100]$ direction, 
because the coupling via the nearest neighbor interaction disappears 
in this case, the dominant coupling is given by the next nearest neighbor 
interactions as 
\begin{eqnarray}
  F_{T100}(\bmp) &=& \Upsilon_{T100} (2 \sin p_x \sin p_y), 
  \label{Eq:FT100}
\end{eqnarray}
while for transverse phonons propagating along the $[110]$ direction, 
\begin{eqnarray}
  F_{T110}(\bmp) &=& \Upsilon_{T110} ( \cos p_x- \cos p_y), 
  \label{Eq:FT110}
\end{eqnarray}
where 
$\Upsilon_{T100}= -g_{\rm n.n.n.}\sqrt{L_z/\rho_{\rm ion} (s_{T100})^2}$ 
[$\Upsilon_{T110}= -g_{\rm n.n}\sqrt{L_z/\rho_{\rm ion} (s_{T110})^2}$] 
with system's $c$-axis dimension $L_z$. 
The electron-phonon vertex function for the longitudinal phonons is given by 
\begin{eqnarray}
  F_{L}(\bmp) &=& \Upsilon_{L} 
  ( \widehat{q}_x^2 \cos p_x +  \widehat{q}_y^2 \cos p_y) 
  \label{Eq:FL} 
\end{eqnarray}
with 
$\Upsilon_{L}= -g_{\rm n.n.}\sqrt{L_z/\rho_{\rm ion} (s_{L})^2}$. 
The important point here is that in each case 
$F_{\{\widehat{\bmK},\nu\}}(\bmp)$ can be expressed as 
$F_{\{\widehat{\bmK},\nu\}}(\bmp) = \Upsilon \, W_\bmp$ 
with a form factor $W_\bmp$ and the corresponding 
electron-phonon coupling constants $\Upsilon$. 
For $T100$ phonons, $W_\bmp$ is equal to the $d_{xy}$-symmetry form factor 
$d'_\bmp = 2 \sin p_x \sin p_y $, 
while it is equal to the $d_{x^2-y^2}$-wave form factor 
$d_\bmp$ for $T110$ phonons. 
In case of longitudinal sounds, $W_\bmp= s_\bmp+ d_\bmp$ for $L100$ phonons 
while $W_\bmp= s_\bmp$ for $L110$ phonons 
where $s_\bmp$ is defined below Eq.~(\ref{Eq:HV05}). 
We note that the inclusion of an isotropic vertex function 
into Eq.~(\ref{Eq:FL}), which ensures the charge neutral condition~\cite{Walker01} 
for longitudinal phonons, does not change the main result discussed below.

Using the electron-phonon interaction (\ref{Eq:H_el-ph}), 
the sound attenuation is calculated from 
the phonon Green's function 
\begin{eqnarray} 
  [D_\nu(\bmk, \ui \Om)]^{-1} &=& [D_\nu^{(0)}(\bmk,\ui \Om)]^{-1}- 
  \Pi_\nu(\bmk,\ui \Om), 
\end{eqnarray}
where 
$  D_\nu^{(0)}(\bmk, \ui \Om) = -(\omega^{(0)}_{\bmk,\nu})^2/ 
  [{\Om^2 + (\omega^{(0)}_{\bmk,\nu})^2}]$
is the bare phonon Green's function, and 
$\Pi_\nu(\bmk,\ui \Om)$ is the phonon self-energy. 
The retarded phonon self-energy, 
$\Pi_\nu^R(\bmk, \ui \Om ) = \Pi_\nu(\bmk, \ui \Om \to \Omega + \ui 0_+)$, 
determines the phonon lifetime as 
$1/\tau_{{\rm ph},\nu}= -(\omega^{(0)}_{\bmk,\nu})^2 \frac{1}{\Omega} {\rm Im} 
\Pi_\nu^R(\bmk, \Omega)$, and the sound attenuation 
$\alpha_{\nu}(\bmk)$ for the phonon with wave vector $\bmk$ and 
polarization $\nu$ is given by 
\begin{eqnarray}
  \alpha_{\nu}(\bmk) &=&  
  -\left( \frac{(\omega^{(0)}_{\bmk,\nu})^2}{s^{(0)}_\nu} \right) 
  \frac{1}{\Omega}{\rm Im} \Pi^{R}_{\nu}({\bmk},\Omega). 
  \label{Eq:alphaDEF}
\end{eqnarray}
Strictly speaking, the sound velocity in Eq.~(\ref{Eq:alphaDEF}) should 
be understood as a renormalized one, 
\begin{eqnarray}
s_\nu &=& s^{(0)}_\nu \sqrt{1- {\rm Re}\Pi^{R}_{\nu}({\bmk},\Omega \to 0)}, 
\end{eqnarray}
with the bare velocity $s^{(0)}_\nu$. 
However, because the renormalization of the sound velocity 
due to the collective fluctuation is usually quite small 
in the experimental resolution~\cite{Luethi70} 
[$(s_\nu-s^{(0)}_\nu)/s^{(0)}_\nu \approx 10^{-3}-10^{-4}$], 
we use the bare sound velocity in Eq.~(\ref{Eq:alphaDEF}). 
Note however that the unrenormalized sound velocity
should be used only in Eq.~(\ref{Eq:alphaDEF}).
As mentioned at the end of Sec.~II, the measurement of the 
softening of the sound velocity is also a key experiment to detect 
the $d_{x^2-y^2}$-wave PI. 

\begin{figure}[t] 
\scalebox{0.7}[0.7]{\includegraphics{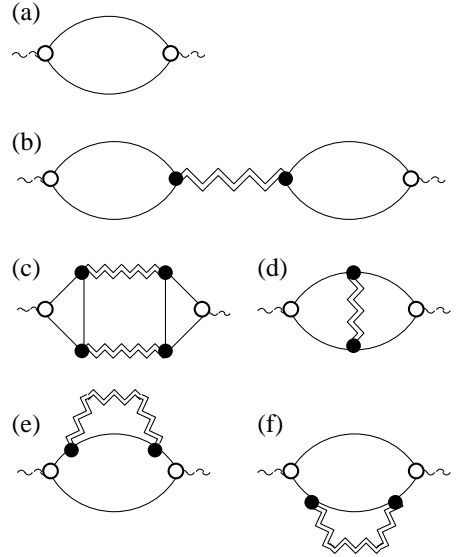}}
\caption{Diagrams for the phonon self-energy $\Pi_\nu(\bmk,\ui \Om)$. 
An open circle represents the electron-phonon vertex 
in Eq.~(\ref{Eq:Vertex_el-ph}), 
a full circle the $d_{x^2-y^2}$-wave vertex $d_\bmp$, 
a wavy line the phonon Green's function, and a double zigzag line 
the fluctuation propagator defined in FIG.~\ref{Fig:fluct_diag01}.}
\label{Fig:Pi01}
\end{figure}

Consider first the sound attenuation caused by itinerant electrons 
in the absence of the $d_{x^2-y^2}$-wave PI. In this case the irreducible 
phonon self-energy is given by~\cite{AGD} [FIG.~\ref{Fig:Pi01}(a)] 
\begin{eqnarray}
  \Pi^{(0)}_\nu(\bmk, \ui \Om) &=& \Upsilon^2 \; T \sum_{\en} \int_{\bmp} 
  W_\bmp^2 \; G_{\sigma^*}^{(0)}(\bmp,\ui \en)  \nonumber \\
  && \hspace{1cm} \times G_{\sigma^*}^{(0)}(\bmp+\bmk ,\ui \en + \ui \Om). 
  \label{Eq:Pi0}
\end{eqnarray}
This gives the sound attenuation without the Fermi surface 
fluctuations, $\alpha^{\rm MF}_{\nu}(\bmk) = 
-[ (\omega^{(0)}_{\bmk,\nu})^2/s_\nu ] 
  \frac{1}{\Omega}{\rm Im} \Pi^{(0)R}_{\nu}({\bmk},\Omega)$.

Now we consider the effect of the $d_{x^2-y^2}$-wave PI. 
The relevant diagrams are shown in FIG.~\ref{Fig:Pi01}(b)-(f) 
where we have neglected impurity vertex corrections due to 
the reason mentioned above Eq.~(\ref{Eq:DysonEQ01}). 
The simplest diagram is given by FIG.~\ref{Fig:Pi01}(b) 
which corresponds to the sound attenuation 
derived from the mode-coupling term~(\ref{Eq:dFGL1a}).  
Diagram (b) gives the sound attenuation derived 
from the mode-coupling term (\ref{Eq:dFGL2a}), and is analogous to 
the Aslamazov-Larkin diagram~\cite{Aslamazov} for the fluctuation conductivity 
in a superconductor. Here we have neglected the vertex 
corrections due to mode-coupling 
[e.g., FIG.~1(c) in Ref.~\onlinecite{Ramazashvili}]. 
Diagrams (d)-(f) give the contributions 
which are not described by the phenomenological (bosonic) 
argument in Sec.~II, and analogous to the Maki-Thompson 
diagram~\cite{Maki68,Thompson70} 
[diagram (d)] and density of states diagrams~\cite{Abrahams70} 
[diagram (e) and (f)]. Note however that the analytical expression 
of each diagram is different from that of a superconducting fluctuation 
contribution because the PI is a kind of diagonal long range order while the 
superconductivity is an off-diagonal long range order. 

For the moment, we focus on the transverse sound attenuation 
because it is this case where 
the fluctuation sound attenuation is propagation-direction 
selective. 
We first consider diagram (b). 
For the transverse sound along $[100]$ 
direction, 
because the electron-phonon vertex function $F_{T100}(\bmp)$ 
[Eq.~(\ref{Eq:FT100})] has the $d_{xy}$-symmetry, 
the phonon self-energy does not couple to the collective fluctuation 
of the $d_{x^2-y^2}$-wave PI. 
Hence the phonon self-energy has no divergent contribution, and is given by 
\begin{eqnarray}
  \Pi_{T100}(\bmk, \ui \Om) &=& \Pi^{(0)}_{T100}(\bmk, \ui \Om), 
\end{eqnarray}
where $\Pi^{(0)}_{T100}$ is defined in Eq.~(\ref{Eq:Pi0}) with 
$W_\bmp= d'_\bmp$. 
On the other hand, the transverse sound along $[110]$ direction 
does couple to the $d_{x^2-y^2}$-wave Fermi surface fluctuation because 
the electron-phonon vertex function $F_{T110}(\bmp)$ 
is proportional to the $d_{x^2-y^2}$-wave form factor. This gives 
the divergent phonon self-energy 
\begin{eqnarray}
  \Pi_{T110}(\bmk, \ui \Om) &=& -(\Upsilon_{T110})^2 \chi_d(\bmk, \ui \Om), 
  \label{Eq:PiT110}
\end{eqnarray}
where $\chi_d(\bmk, \ui \Om)$ behaves as Eq.~(\ref{Eq:chid01}) 
near the second order transition. 

\begin{figure}[t] 
  \begin{center}
    \scalebox{0.5}[0.5]{\includegraphics{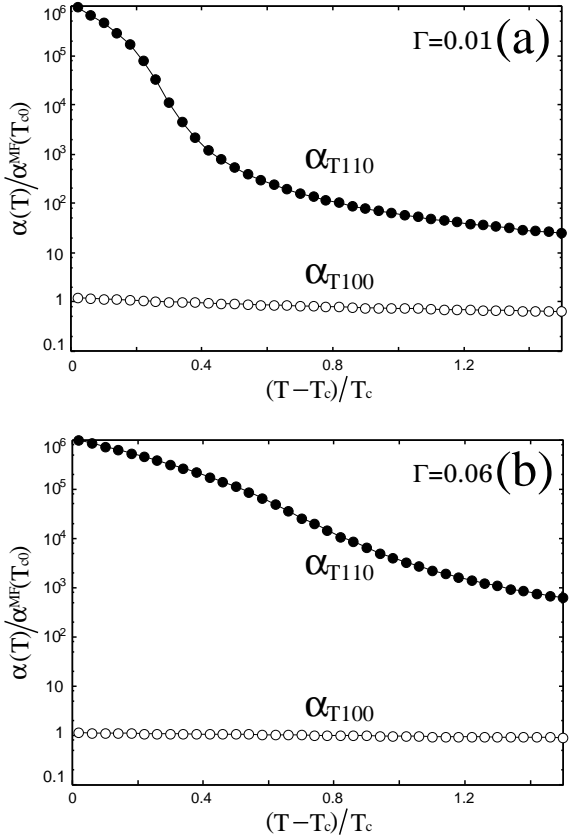}}
  \end{center}
\caption{Calculated ultrasound attenuation for $T100$ phonons (open circles) 
and $T110$ phonons (full circles) for $\Gamma=0.01$ [(a)] and 
$\Gamma=0.06$ [(b)] at constant field $h=h_{\rm VH}$. 
The parameters used in this figure are the same 
as in FIG.~\ref{Fig:MFPD} and FIG.~\ref{Fig:alpR-temp}. 
The data are normalized by $\alpha^{\rm MF}$ 
at the mean field transition temperature $T_{c0}$ where 
$\alpha^{\rm MF}$ is defined below Eq.~(\ref{Eq:Pi0}). } 
\label{Fig:aT100aT110-dt}
\end{figure}

\begin{figure}[t] 
  \begin{center}
    \scalebox{0.49}[0.49]{\includegraphics{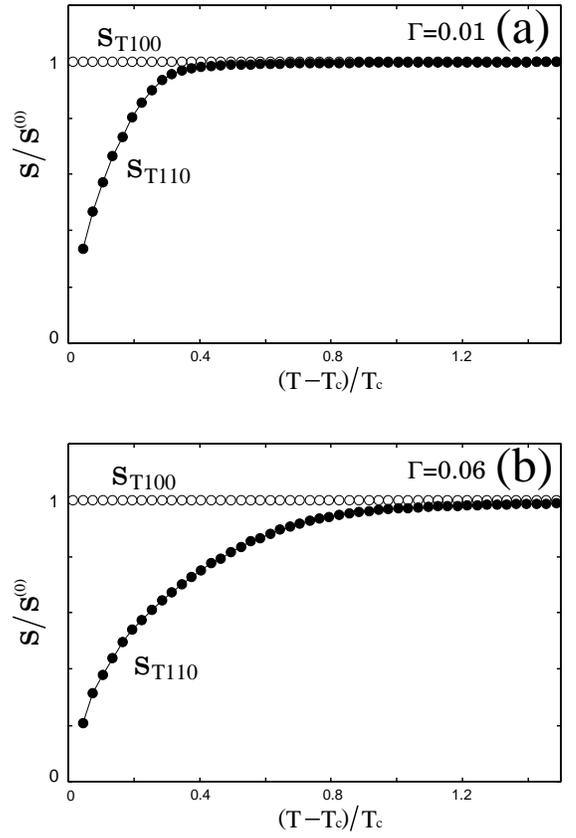}} 
  \end{center}
\caption{Calculated ultrasound velocity for $T100$ phonons (open circles) 
and $T110$ phonons (full circles) for $\Gamma=0.01$ [(a)] and 
$\Gamma=0.06$ [(b)] at constant field $h=h_{\rm VH}$. 
The parameters used in this figure are the same 
as in FIG.~\ref{Fig:MFPD} and FIG.~\ref{Fig:alpR-temp}. 
The data are normalized by the bare sound velocity $s^{(0)}$. 
Here, we have used a moderately 
large electron-phonon coupling constant 
$\Upsilon^2=0.001$ to enlarge the signal, whereas the actual softening 
would be much smaller.~\cite{Luethi70} 
}
\label{Fig:vsT110T100-dt}
\end{figure}

Next we consider diagram (c). In this case, it is easy to show 
that the triangle blocks in this diagram vanish for both 
$T100$ and $T110$ phonons, hence there is no contribution 
from this diagram. This is consistent with the argument 
in Sec.II that Eq.~(\ref{Eq:dFGL2a}) couples only to the 
longitudinal sound, because diagram (b) gives the sound attenuation 
derived from this mode-coupling term (\ref{Eq:dFGL2a}). 

We then consider diagrams (d)-(f) which were not taken into 
account in the phenomenological argument in Sec.~II. 
Note that if we pick up only the contribution 
just at the Fermi surface assuming a constant density of states, 
then we can show that diagram (d) and diagrams (e) and (f) cancel. 
However, once we include the energy dependence of the density of states, 
there is a nonzero contribution from these diagrams 
as in the case of the quartic term of the GL functional. 
By power counting, we can show that these diagrams give a 
logarithmically divergent sound attenuation 
$\alpha_T \sim \ln \mu$ where $\mu$ is the renormalized mass term 
(\ref{Eq:deltaSCR}). We think that in the actual experiment 
this logarithmic divergence is negligible compared with the 
stronger divergence caused by Eq.~(\ref{Eq:PiT110}). 

Now we come to the case with longitudinal sound modes. 
Concerning the fluctuation contribution to the longitudinal sound attenuation, 
it is sufficient to consider diagram (c) in FIG.~\ref{Fig:Pi01} 
irrespective of the propagation direction 
because as you can see in the following this diagram always gives the divergent 
behavior with the same exponent. 
After analytical continuation, diagram (c) gives 
\begin{eqnarray}
  \Pi_L^R(\bmk, \Omega) &\sim& -\ui \Omega {\cal B}^2 \int 
  \frac{d \omega}{\sinh^2(\frac{\omega}{2T})} 
  \int d^3 Q' \Big[ {\rm Im} \chi_d^R(\bmQ',\omega)\Big]^2, 
\label{Eq:ALterm01}
\end{eqnarray}
where 
  ${\cal B} = \Upsilon_L \sum_{\en} 
  \int_\bmp W_\bmp d_\bmp^2 [G^{(0)}(\bmp,\ui \en)]^3 
  = 
  \frac{\Upsilon_L}{2} 
  \int_\bmp W_\bmp d_\bmp^2 
  f''_\Gamma(\xi_{\bmp,\sigma^*}) $
comes from the triangle block in the diagram and  
$\bmQ' = (\bmq, Q_z/\gamma)$ is the rescaled three-dimensional 
wavevector introduced in Eq.~(\ref{Eq:Qaniso}). 
Roughly speaking, ${\cal B}$ is proportional to the degree of 
particle-hole asymmetry, 
and in case of a perfect particle-hole symmetry this contribution vanishes. 
By power counting, we can show that this gives a 
divergent sound attenuation 
$\alpha_L \sim \mu^{-2}$ where $\mu$ is the renormalized mass term 
(\ref{Eq:deltaSCR}). 
Therefore, as was already discussed in Sec.~II, 
the longitudinal sound attenuation always show a divergent behavior 
on approaching a second order PI.

To summarize, a microscopic calculation of the 
fluctuation sound attenuation in this subsection 
gives the same result as obtained by a phenomenological argument 
in Sec.~II. 
Results of this subsection coincide 
with the main results in Sec.~II, i.e., 
Eqs.~(\ref{Eq:alpT110}), (\ref{Eq:alpT100}) and (\ref{Eq:alpLall}), 
if we replace the bare mass $\mu_0$ with the renormalized mass $\mu$. 

In FIG.\ref{Fig:aT100aT110-dt} we plot the transverse sound attenuation 
along $[100]$ and $[110]$ directions 
as functions of temperature. 
We also plot in FIG.\ref{Fig:vsT110T100-dt} the transverse sound velocities 
along $[100]$ and $[110]$ directions as functions of temperature. 
As was already discussed in Sec.~II, there is a fluctuation contribution 
to $[110]$ phonons while there is no contribution to $[110]$ phonons. 
Comparing data for a cleaner system [FIG.\ref{Fig:aT100aT110-dt}(a)] 
to that for a dirtier system [FIG.\ref{Fig:aT100aT110-dt}(b)], we see 
that the dirtier system shows broader fluctuation behavior 
in the reduced temperature $(T-T_c)/T_c$. 

\section{Discussion and Conclusion} 
We have studied 
the effect on sound properties 
of the Fermi surface fluctuations near a $d_{x^2-y^2}$-wave PI, 
and discussed that there is a propagation-direction-dependent selection rule in the 
fluctuation {\it transverse} sound attenuation and sound velocity softening. 
As is shown in FIG.~\ref{Fig:uLuT}, FIG.~\ref{Fig:aT100aT110-dt}, and 
FIG.~\ref{Fig:vsT110T100-dt}, 
the transverse sound attenuation and sound velocity softening 
along $[110]$ direction are enhanced by the Fermi surface fluctuations 
while those along $[100]$ direction are not affected. 
Also it was argued that there are always fluctuation contributions 
to the longitudinal sound attenuation and sound velocity softening. 
We note that a qualitatively similar conclusion can be reached for a second order 
structural transition which breaks the lattice symmetry in the same way, 
but is {\it not} caused by a Fermi surface instability. 
Such a transition would not be identified as a genuine Pomeranchuk instability.

As was already mentioned in Sec.~I, the possibility of the 
$d_{x^2-y^2}$-wave PI discussed in this paper has been
debated~\cite{Grigera04,Kee05,Honerkamp05,Borzi07,Yamase07,Doh07,Puetter07} 
as a possible explanation for the anomalous phase found in 
Sr$_3$Ru$_2$O$_7$ under strong magnetic fields, 
and it provides us with a good opportunity to apply our results. 
In earlier experiments for this material, 
a metamagnetic transition was found~\cite{Perry01,Grigera01} 
at a magnetic field $B_m$ where the magnetization 
shows a steep jump. Since several non-Fermi-liquid properties are 
observed around $B_m$, this system was originally discussed in terms of 
the metamagnetic quantum critical point,~\cite{Millis02} 
and the importance of van Hove singularity for the metamagnetic transition 
was discussed.~\cite{Binz04} 
However, the subsequent experiments using an ultra-pure sample 
found at least two transitions,~\cite{Ohmichi03,Perry04} 
and later on it was revealed~\cite{Grigera04} that the two transitions 
reflect the boundary of a new ordered phase 
which is accompanied by metamagnetic transitions. 
Based on a consideration on resistivity data, 
it was proposed~\cite{Grigera04} and demonstrated~\cite{Kee05} 
that the observed new phase can be an ordered state with $d_{x^2-y^2}$-wave PI. 

On this background, it is tempting to compare our results with the experiments 
for bilayer ruthenate Sr$_3$Ru$_2$O$_7$. 
Firstly, as was already discussed in Ref.~\onlinecite{Yamase07}, 
the calculated phase diagram [FIG.~\ref{Fig:MFPD} (a)] 
looks quite similar to what was observed in experiments 
(Fig.~3 in Ref.~\onlinecite{Grigera04}). 
Secondly, the earlier experiments can be interpreted as a result of 
the impurity effects which narrow the area of the ordered phase 
[FIG.~\ref{Fig:MFPD} (b)], 
because two closely located transition lines (beyond the 
experimental resolution) would look like 
a single transition line in experiments. 
Thirdly, the non-Fermi-liquid behaviors observed in 
resistivity,~\cite{Grigera01,Perry01,Zhou04} 
specific heat,~\cite{Grigera01,Zhou04,Perry05} 
and thermal expansion~\cite{Gegenwart06} 
can be interpreted as a result of the Fermi surface fluctuations 
near the $d_{x^2-y^2}$-wave PI discussed in Sec.~III B. 
This argument can also be applied to the non-Fermi-liquid behavior observed in 
nuclear relaxation rate,~\cite{Kitagawa05} if 
we take into account the spin-orbit scattering. 
These facts suggest that the $d_{x^2-y^2}$-wave PI is a promising scenario 
for the curious phase observed in this material. 
Hence, it is interesting to apply our main result to Sr$_3$Ru$_2$O$_7$, 
i.e., Eqs.~(\ref{Eq:alpT110}), (\ref{Eq:alpT100}) and (\ref{Eq:alpLall}) 
[note that Eq.~(\ref{Eq:alpL110}) is replaced by Eq.~(\ref{Eq:alpLall})], 
and we propose to measure the propagation-direction resolved transverse 
sound attenuation and sound velocity softening in Sr$_3$Ru$_2$O$_7$, 
since it can determine experimentally the presence or absence of the PI in this material.

Finally we comment on the recent papers~\cite{Raghu09,Lee09} 
on the microscopic mechanism of the $d_{x^2-y^2}$-wave PI for Sr$_3$Ru$_2$O$_7$. 
In these articles, the authors point out the importance 
of the quasi-one-dimensional ruthenium orbitals ($d_{xz}$ and $d_{yz}$) 
for the occurrence of the $d_{x^2-y^2}$-wave PI. 
On the other hand, our microscopic analysis in Sec.~III essentially 
uses quasi-two-dimensional ruthenium orbital $d_{xy}$. 
We point out that while the microscopic analysis 
in Sec.~III needs small modifications~\cite{com1} 
if we start from $d_{xz}$ and $d_{yz}$ orbitals, 
the main conclusion remains unchanged 
that the transverse sound attenuation 
along $[110]$ direction and the longitudinal sound attenuation in {\it all} 
directions are enhanced by the Fermi surface fluctuations 
while the transverse attenuation along $[100]$ direction is not affected. 
This can be inferred from the phenomenological nature of the 
argument given in Sec.~II to arrive at 
Eqs.~(\ref{Eq:alpT110}), (\ref{Eq:alpT100}) and (\ref{Eq:alpLall}).

\acknowledgments 
We are grateful to S. Fujimoto for his help in the early stage of our 
research. One of us (H. A.) would like to thank 
D. Agterberg, B. Binz, K. Ishida, Y. Maeno, and H. Yamase for useful comments, and 
M. Ossadnik, A. R\"{u}egg, and all the member of the condensed matter group 
in ITP at ETH Z\"{u}rich for valuable discussions. 
This study was financially supported through a fellowship of the Japan 
Society for the Promotion of Science and the NCCR MaNEP of the Swiss 
Nationalfonds. A part of the numerical calculations was carried out 
on Altix3700 BX2 at YITP in Kyoto University. 

\appendix 

\section{Calculation of $\chi^{{\rm irr}, R}_{d,0}(\bmq,\omega)$} 

\begin{figure}[t] 
  \begin{center}
    \scalebox{0.75}[0.75]{\includegraphics{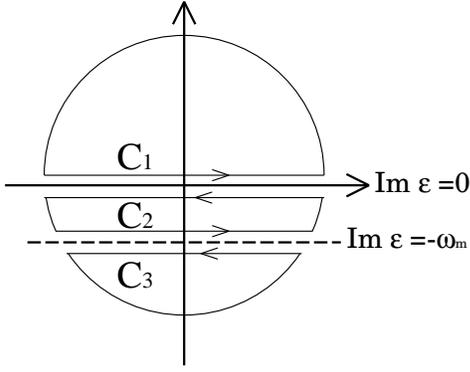}} 
  \end{center}
\caption{Contour to perform the sum over $\en$.} 
\label{Fig:contour01}
\end{figure}

To evaluate $\chi^R_{d,{\rm RPA}}(\bmq,\omega)$, we need to calculate 
$\chi^{{\rm irr},R}_{d,0}(\bmq,\omega)$. 
On Matsubara (imaginary) axis, $\chi^{\rm irr}_{d,0}(\bmq, \ui \om)$ 
is given by Eq.~(\ref{Eq:chi_irr_d0}). 
We transform the Matsubara sum 
into an integral over the contour as shown in FIG.~\ref{Fig:contour01} 
using $T \sum_{\en} F(\ui \en)= \oint \frac{d \varepsilon}{4 \pi \ui} 
\tanh(\frac{\varepsilon}{2T}) F(\varepsilon)$, and 
perform analytical continuation $\ui \om \to \omega + \ui 0_+$. 
Contour ${\cal C}_1$ and ${\cal C}_3$ give the real part of 
$\chi^{R}_{d,0}(\bmq,\omega)$ for small 
$\omega$, and in this case we can evaluate this quantity in the limit $\omega \to 0$, 
\begin{eqnarray}
  {\rm Re} \chi^{{\rm irr}, R}_{d,0}(\bmq,\omega) 
  &=&
  \int \frac{d \varepsilon}{ 4 \pi \ui} 
  \tanh \left( \frac{\varepsilon}{2T} \right)   \int_\bmp d^2_\bmp \nonumber \\
  &\times& 
  \Big\{ \big[ G_{\sigma^*}^{(0)R}(\bmp+\bmq,\varepsilon)
  G_{\sigma^*}^{(0)R}(\bmp,\varepsilon) \big] - c.c. \Big\} \nonumber \\
  &=& 
  T \sum_{\en} 
  \int_\bmp d^2_\bmp 
  \big[ G_{\sigma^*}^{(0)}(\bmp+\bmq,\ui \en) 
  G_{\sigma^*}^{(0)}(\bmp,\ui \en) \big] \nonumber \\
  &=& 
  \int_\bmp d^2_\bmp \bigg\{  f{'}_\Gamma(\xi_{\bmp,\sigma^*}) \nonumber \\
  &+& 
  \Big[   \frac{f{''}_\Gamma(\xi_{\bmp,\sigma^*}) }{4} (M_x +M_y) \Big] \bmq^2 
  \nonumber \\ 
  &+&
  \Big[ \frac{f{'''}_\Gamma(\xi_{\bmp,\sigma^*})}{12} 
  (v_x^2 +v_y^2) \Big] \bmq^2 \bigg\} + O(\bmq^4), 
\end{eqnarray}
where we have used 
$\frac{\partial}{\partial \xi_\bmp} G^{(0)}(\bmp, \ui \en ) = 
[G^{(0)}(\bmp, \ui \en )]^2$ and 
$T \sum_{\en} [ G^{(0)}_{\sigma} (\bmp,\ui \en) ]^{n+1} 
= \frac{1}{n!} \frac{\partial^n } {\partial \xi_{\bmp,\sigma}^n } 
f_\Gamma(\xi_{\bmp,\sigma}) $, 
and $M_j (j+x,y)$ is given below Eq.~(\ref{Eq:w0DEF}). 

On the other hand, contour ${\cal C}_2$ gives the imaginary part of 
$\chi^{{\rm irr},R}_{d,0}(\bmq,\omega)$ for small $\omega \ll v_{\rm F}|\bmq| $ as 
\begin{eqnarray}
  {\rm Im} \chi^{{\rm irr},R}_{d,0}(\bmq,\omega) 
  &=&
  -\int \frac{d \varepsilon}{ 4 \pi \ui} 
  \int_\bmp d^2_\bmp 
  \left( \frac{\tanh \left( \frac{\varepsilon}{2T} \right)}
  {\ui/\tau- {\bm v}_p \cdot \bmq} \right)
  \nonumber \\ 
  &\times& 
  \frac{\partial}{\partial \varepsilon} 
  \big[ G_{\sigma^*}^{(0)R}(\bmp,\varepsilon) - c.c. \big] \omega 
  + O(\omega^2) \nonumber \\
  &=& 
  -T \sum_{\en} \int_\bmp 
  d^2_\bmp 
  \left( \frac{\ui \omega}
    {2 \Gamma+ \ui {\bm v}_p \cdot \bmq} \right) \nonumber \\ 
  && \hspace{1cm} \times
  \frac{\partial}{\partial \xi_\bmp} 
  G_{\sigma^*}^{(0)}(\bmp,\ui \en) + O(\omega^2) \nonumber \\
  &=& 
  \frac{\ui \omega}{|\bmq|} 
  \int_\bmp d^2_\bmp 
  \left( \frac{-f{'}_\Gamma(\xi_{\bmp,\sigma^*})}
    {2 \Gamma/|\bmq|+ \ui {\bm v}_p \cdot \widehat{\bmq}} \right)
  + O(\omega^2), 
\end{eqnarray}
where we have used 
$\frac{\partial}{\partial \varepsilon} G_\sigma^{(0)R/A}(\bmp,\varepsilon) = 
- \frac{\partial}{\partial \xi_\bmp} G_\sigma^{(0)R/A}(\bmp,\varepsilon)$.


\begin{thebibliography}{99}
\bibitem{com1}
  The form of the electron-phonon vertex should be changed 
  if we start from $d_{xz}$ and $d_{yz}$ orbitals, but this does not 
  change our conclusion.
\end{thebibliography}

\end{document}